\documentclass[12pt]{article}

\usepackage{amsmath,amsthm,amssymb,amsfonts,graphics,graphicx,amscd,amsfonts,epsf,epsfig,color}

\makeatletter
 
  \@addtoreset{equation}{section}
 \makeatother

\newcommand{\id}{{1\!\!1}} 
\newcommand {\be}{\begin{equation}}
\newcommand {\ee}{\end{equation}}
\newcommand {\bea}{\begin{eqnarray}}
\newcommand {\eea}{\end{eqnarray}}
\newcommand {\nn}{\nonumber}
\newcommand {\tr}{{\rm tr\,}}

\newcommand {\cN}{{\cal N}}

\newcommand {\cM}{{\cal M}}
\newcommand {\cO}{{\cal O}}
\newcommand {\cR}{{\cal R}}
\newcommand {\cS}{{\cal S}}

\newcommand{\cD}{{\cal D}}

\newcommand{\Z}{{\mathbb Z}}
\newcommand{\C}{{\mathbb C}}
\newcommand{\del}{\partial}
\newcommand{\e}{{\rm e}}

\newtheorem*{lemma}{Lemma}
\newtheorem{lem}{Lemma}

\allowdisplaybreaks[4]

\date{}
\begin{document}
\thispagestyle{empty} \addtocounter{page}{-1}
\begin{flushright} 
YITP-17-57
\end{flushright} 

\vspace{0.1cm}

\begin{center}
  {\large \bf
  
${\cal O}(a)$ Improvement of 2D ${\cal N}=(2,2)$ Lattice SYM Theory\\
  }
\end{center}
\vspace{0.1cm}
\vspace{0.1cm}
\begin{center}

Masanori Hanada$^{a,b,c}$\footnote{masanori.hanada@colorado.edu},  
Daisuke Kadoh$^{d}$\footnote{kadoh@keio.jp}, 
So Matsuura$^d$\footnote{s.matsu@phys-h.keio.ac.jp}
and Fumihiko Sugino$^e$\footnote{fusugino@gmail.com}
 
\vspace{0.3cm}

$^a${\em \small Yukawa Institute for Theoretical Physics, Kyoto University, \\
Kitashirakawa Oiwakecho, Sakyo-ku, Kyoto 606-8502, Japan}

$^b${\em \small The Hakubi Center for Advanced Research, Kyoto University, \\
Yoshida Ushinomiyacho, Sakyo-ku, Kyoto 606-8501, Japan}

$^c${\em \small Department of Physics, University of Colorado, Boulder, CO 80309, USA}

$^d${\em \small Hiyoshi Departments of Physics, and Research and Education Center for Natural Sciences, \\
Keio University, 4-1-1 Hiyoshi, Yokohama, Kanagawa 223-8521, Japan}

$^e${\em \small Center for Theoretical Physics of the Universe, \\
Institute for Basic Science (IBS), Seoul 08826, Republic of Korea}

\end{center}

\vspace{1.5cm}

\begin{center}
  {\bf abstract}
\end{center}

We perform a tree-level ${\cal O}(a)$ improvement of two-dimensional 
${\cal N}=(2,2)$ supersymmetric Yang-Mills theory on the lattice, 
motivated by the fast convergence in numerical simulations. 
The improvement respects an exact supersymmetry $Q$ 
which is needed for obtaining the correct continuum limit without a parameter fine tuning.  
The improved lattice action is given within a milder locality condition 
in which the interactions are decaying as the exponential of the distance on the lattice. 
We also prove that the path-integral measure is invariant under the improved $Q$-transformation.

\newpage
\tableofcontents
\section{Introduction}
Nonperturbative dynamics of supersymmetric gauge theories is of great interest
for various reasons. 
However many interesting problems are out of reach of the current analytic understandings.
For example, while properties of the vacua can be analyzed very precisely 
by using holomorphy (e.g.~\cite{Seiberg:1994rs,Seiberg:1994pq}), 
the dynamics of excitations is hard to study except for specific objects saturating the so-called BPS bound. 
In the context of the gauge/gravity duality, gauge theory dual to weakly coupled gravity 
is strongly coupled and furthermore the quantities of interests are not necessarily 
protected by supersymmetry. 
Numerical simulation based on lattice gauge theory is considered as a promising approach to such problems.

Historically, lattice formulation of supersymmetry brushed off many attempts. 
The problem was the {\it parameter fine tuning problem}; because a lattice breaks 
the infinitesimal translation, which is a part of the supersymmetry algebra, the supersymmetry 
cannot be preserved completely on a lattice. Then, even if the supersymmetry is restored in 
a naive continuum limit at the tree level, radiative corrections break it in general.  

For two-dimensional theories, the parameter fine tuning problems can be circumvented 
by keeping a part of the supersymmetry algebra (one or two supercharges and 
$U(1)$ or $SU(2)$ R-symmetry) at discretized level 
\cite{
Kaplan:2002wv,
Catterall:2003wd,
Cohen:2003xe,
Cohen:2003qw,
Sugino:2003yb, 
Sugino:2004qd, 
DAdda:2004jb,
Sugino:2004uv, 
Kaplan:2005ta,
Sugino:2006uf, 
Endres:2006ic,
Giedt:2006dd,
Catterall:2007kn,
Matsuura:2008cfa,
Sugino:2008yp,
Kikukawa:2008xw,
Kadoh:2009yf,
Kanamori:2012et}.
Encouraged by this development, several groups have been trying lattice simulations of 
two-dimensional super Yang-Mills theories \cite{Kanamori:2008bk,Kanamori:2008yy,Hanada:2009hq,Hanada:2010qg,Catterall:2011aa},\footnote{
Another approach with a parameter fine tuning can be found in Refs.~\cite{Suzuki:2005dx,August:2016orf,August:2017eox}. Also, for four-dimensional $\cN=1$ super Yang-Mills theory, see \cite{Montvay:1995ea,Montvay:2001aj,Bergner:2015adz,Ali:2017ijx}.
} 
including the maximally supersymmetric theory relevant for the gauge/gravity duality 
\cite{Catterall:2010fx,Giguere:2015cga,Kadoh:2017mcj,Catterall:2017lub}. 
Furthermore it has been pointed out that such two-dimensional theories
can be used to construct four-dimensional super Yang-Mills theory~\cite{Hanada:2010kt,Hanada:2010gs}
by utilizing the Myers effect with which two spatial dimensions are encoded in matrix degrees of freedom~\cite{Myers:1999ps,Maldacena:2002rb}.  
The two-dimensional $\cN=(2,2)$ supersymmetric Yang-Mills theory on an arbitrarily discretized Riemann surface is developed in 
\cite{Matsuura:2014kha,Matsuura:2014nga, Kadoh:2016eju, Kamata:2016xmu, Kamata:2016rqi}.

At present, there is a consensus among the community of researchers 
that these regularization schemes can work in principle. 
However, in practice --- especially, in order to perform precision measurements 
at large volume and large $N$, with currently available numerical resources --- 
it is desirable to have improved regularizations which converge to the continuum limit faster. 
In the lattice QCD community, such improvement is known as the Symanzik improvement program~\cite{Symanzik:1979ph,Weisz:2010nr}. 
Errors arising on discretization of a continuum system by lattice are of the order of the lattice spacing $\cO(a)$ in general. 
The program makes the errors reduced to higher orders in $a$. 
  
In this paper, we consider the improvement of the lattice action of 
two-dimensional ${\cal N}=(2,2)$ super Yang-Mills theory proposed by one of the authors (F.~S.) \cite{Sugino:2003yb}. 
This is a technically nontrivial subject, because the improvement term must preserve the exact 
supercharge and R-symmetry that are relevant to realize the theory flowing to the desired continuum theory without 
any parameter fine tuning.  
As a first step, we consider ${\cal O}(a)$ improvement at the tree level. 
Because the ultraviolet divergence is mild due to the low-dimensionality and supersymmetry, 
it is sufficient to consider the tree and one-loop levels in this setting.~\footnote{
As a comparison, the tree-level improvement is enough to achieve the significant acceleration of the simulation 
in the case of the $(0+1)$-dimensional theory~\cite{Berkowitz:2016jlq}.
}  
The one-loop calculation will be reported in the forthcoming publication.

The previous unimproved action \cite{Sugino:2003yb} has been constructed in the three steps:
\begin{itemize}
\item
Write the action $S$ in the continuum theory in a $Q$-exact form, $S=Q \frac{1}{2g^2} \int d^2x\, \Xi(x)$, 
by using one of the supercharges $Q$. 

\item
Construct a lattice counterpart of $Q$-transformation, $Q_{\rm lat}$, which is an exact symmetry at the regularized level. 

\item
Define a lattice action in the  $Q$-exact form as $S_{\rm lat}=Q_{\rm lat} \frac{a^2}{2g^2} \sum_x \Xi_{\rm lat}(x)$
where $\Xi_{\rm lat}$ is a lattice counterpart of $\Xi(x)$.  

\end{itemize}
Note that the path-integral measure has to be taken in a $Q$-invariant way as well, and the natural measure is actually $Q$-invariant. 
The created lattice action reproduces the continuum one with ${\cal O}(a)$ corrections
because the lattice supersymmetry transformation 
generated by 
$Q_{\rm lat}$ and the term $\Xi_{\rm lat}$ are different from the continuum ones at the order.

For ${\cal O}(a)$ improvement, we need to improve both of 
$Q_{\rm lat}$ and $\Xi_{\rm lat.}$, 
keeping the path-integral measure $Q$-invariant. 
Note also that we have to make sure that an extra ${\cal O}(a)$ correction does not appear from the measure. 
The improvements will be done by lattice operators $\cR_\mu$ and $\cR_{12}$ for $Q$ and $\Xi$, respectively, 
which include a kind of Wilson terms. 
We will find that the improved lattice theory satisfies a milder locality condition known as the exponential locality. 
Such condition is accepted in obtaining a local continuum theory within the universality hypothesis~\cite{Wilson:1973jj,Niedermayer:1998bi,Hernandez:1998et}.    

This paper is organized as follows. In Sec.~\ref{sec:original theory} we review
the continuum theory and introduce the unimproved lattice action. 
On the way we clarify the origin of ${\cal O}(a)$ deviation from the continuum theory to illuminate 
a strategy for the improvement. 
Sec.~\ref{sec:improvement} and  Sec.~\ref{sec:final result}
are the main parts of this paper. 
In Sec.~\ref{sec:improvement}, 
clarifying existence conditions of the consistent lattice $Q$-transformation by a lemma,  
we improve the lattice action with keeping the $Q$-exact form. 
In  Sec.~\ref{sec:final result}, 
we summarize the improved theory and show that 
it is free from the doubling problem thanks to $\cO(a^3)$ Wilson terms.
The $Q$-invariant measure is then consistently defined in any finite physical volume.  
Sec.~\ref{sec:conclusion} is devoted to summarize the results and discuss future directions 
concerning this project. 
A proof of the lemma is given in appendix~\ref{app:lemma}, 
and the factors $\cR_\mu$ and $\cR_{12}$ are presented with their locality properties in 
appendix~\ref{app:R}. 
For convenience in the numerical simulation, we present the explicit form of the improved lattice action in appendix~\ref{app:action}. 
Appendix~\ref{app:determinant} gives a computational detail related to the $Q$-invariance of 
the path-integral measure.   

\section{Original lattice formulation}\label{sec:original theory}
In this section, we briefly review two-dimensional $\cN=(2,2)$ supersymmetric Yang-Mills theory and its 
lattice formulation with one of the four supercharges of the theory exactly preserved. As a preparation to the $\cO(a)$ improvement, 
errors arising in the discretization are also discussed.   

\subsection{Continuum theory}\label{subsec:continuum}
We start with 
$\cN=(2,2)$ supersymmetric Yang-Mills theory on two-dimensional Euclidean space, 
whose field contents are gauge fields $A_\mu$, adjoint scalars $\phi$, $\bar{\phi}$, 
gaugino fields $\lambda$, $\bar{\lambda}$ and an auxiliary field $D$ which satisfies $D^\dag=-D$. 
  We assume that the gauge group is $SU(N)$ and all fields are matrix-valued functions which are expanded by a basis of $N\times N$ traceless hermitian matrices $T_\alpha$ ($\alpha=1,\cdots, N^2-1$) normalized as $\tr (T_\alpha T_\beta)=\delta_{\alpha\beta}$. 

The action is expressed as 
\bea
S_{\rm cont} & = & \frac{1}{2g^2}\int d^2x\,\tr\biggl\{\frac12  \sum_{\mu,\nu=1}^2 F_{\mu\nu}^2+  \sum_{\mu=1}^2 D_\mu \phi D_\mu\bar{\phi} +\frac14[\phi,\,\bar\phi]^2-D^2 
\nn \\
  & & +2\bar{\lambda}_R(D_1-i D_2)\lambda_R +2\bar{\lambda}_L (D_1+iD_2) \lambda_L +2\bar{\lambda}_R[\bar{\phi},\lambda_L] 
  +2\bar{\lambda}_L[\phi,\lambda_R]\biggr\} ,
  \label{action_cont}
\eea  
where $g$ is a coupling constant, $\mu, \nu=1,2$, and the subscripts $L$ and $R$ denote the spinor indices of $\lambda$. 
The theory is obtained from the four-dimensional $\cN=1$ supersymmetric Yang-Mills theory by the dimensional reduction. 
In fact, the gaugino fields with/without bars signify four-dimensional chirality and the indices $L$ and $R$ two-dimensional 
chirality~\cite{Sugino:2008yp}. 
The covariant derivatives and the field strengths are defined by 
\begin{eqnarray}
&& D_\mu = \del_\mu + i[A_\mu,\, \cdot \, ], 
\label{covariant_derivative}
\\
&& F_{\mu\nu}=\partial_\mu A_\nu-\partial_\nu A_\mu +i[A_\mu, \,A_\nu],
\label{field_tensor}
\end{eqnarray}
respectively. The action is invariant under an infinitesimal gauge transformation  
with a function $\omega(x)=\sum_\alpha\omega_\alpha(x)\,T_\alpha$:
\begin{eqnarray}
\delta_\omega A_\mu(x) =-D_\mu \omega (x), \qquad \delta_\omega F(x) =i[\omega(x),F(x)], 
\label{gauge_transformation}
\end{eqnarray}
where the adjoint scalar and fermion fields are represented as $F$.
The theory possesses four supersymmetries corresponding to the four spinor components (with/without bars, and $L/R$).  
In addition, there are two types (vector and axial) of the $U(1)_R$ symmetries in the theory. 
Among them, the lattice formulation given in section~\ref{subsec:unimproved lattice} preserves the latter, which 
transforms the fields according to the charge assignment: $+2$ $(-2)$ to $\phi$ $(\bar{\phi})$, $+1$ $(-1)$ to 
$\lambda_L$, $\bar{\lambda}_R$ ($\lambda_R$, $\bar{\lambda}_L$), and 0 to the others. 

Naive lattice discretization breaks supersymmetry completely, 
due to the lack of Leibniz rule on the lattice. 
This difficulty can be partly avoided by expressing the action as an exact form with respect to a nilpotent supercharge $Q$.
Namely, one of the four supercharges, $Q$, can be kept on a lattice.  
To find the exact form of the action, it is convenient to use new fermion variables $\psi_\mu, \chi, \eta$ defined as
\bea
&& \psi_1\equiv \frac{1}{\sqrt{2}}(\lambda_L+\bar{\lambda}_R),  \qquad
 \psi_2\equiv \frac{i}{\sqrt{2}}(\lambda_L-\bar{\lambda}_R), \nn \\
&& \chi\equiv \frac{1}{\sqrt{2}}(\lambda_R-\bar{\lambda}_L), \qquad
 \eta\equiv -i\sqrt{2}(\lambda_R+\bar{\lambda}_L),
\label{rename}
\eea
instead of the original gaugino fields $\lambda, \bar\lambda$.~\footnote{
These variables are used in topological field theory, and here we employ them to define the lattice action with an exact supercharge. The new auxiliary field $H$ is given by $H=iD+iF_{12}$. 
Note $D$ should be taken as anti-hermitian in (\ref{action_cont}). 
Among the four supercharges of the theory ($Q_L$, $Q_R$, $\bar{Q}_L$, $\bar{Q}_R$), a combination $Q\equiv -(Q_L+\bar{Q}_R)/\sqrt{2}$ 
transforms fields as (\ref{cont_QA})--(\ref{cont_Qchi}).
}
The action (\ref{action_cont}) can be expressed in the variables as 
\begin{eqnarray}
&& \hspace{-1cm}
S_{\rm cont} = \frac{1}{2g^2}\int d^2x\,\tr\biggl\{ H^2 -2i HF_{12} +\sum_{\mu=1}^2 D_\mu \phi D_\mu\bar{\phi} +\frac14[\phi,\,\bar\phi]^2 \nn \\
  &&\hspace{0.2cm}  +2i \chi (D_1 \psi_2-D_2 \psi_1) + i \sum_{\mu=1}^2 \psi_\mu D_\mu \eta -\frac14 \eta [\phi,\eta] 
  -\chi[\phi,\chi] +\sum_{\mu=1}^2 \psi_\mu [\bar\phi,\psi_\mu]
    \biggr\}.
  \label{twisted_action_cont}
\end{eqnarray}
We choose one of the  four supercharges $Q$ defined by 
\bea
& & QA_\mu=\psi_\mu, \label{cont_QA} \\
& & Q\psi_\mu=iD_\mu\phi, \label{cont_Qpsi} \\
& & Q\phi=0,  \label{cont_Qphi}  \\
& & Q\bar{\phi}=\eta, \qquad Q\eta=[\phi,\,\bar{\phi}], \label{cont_Qeta} \\
& & Q\chi=H, \qquad QH=[\phi,\,\chi]. \label{cont_Qchi}
\eea
From (\ref{gauge_transformation}) and (\ref{cont_QA})-(\ref{cont_Qchi}), 
we can see the nilpotency of $Q$ up to an infinitesimal gauge transformation with the parameter $\phi$,
\bea
Q^2 =-i\delta_\phi. 
\label{Q_nilpotency_cont}
\eea
By using this $Q$, 
the action can also be recast as a $Q$-exact form~\cite{Witten:1988ze,Witten:1990bs}: 
\be
S_{\rm cont} =Q\,\frac{1}{2g^2}\int d^2x\,\Xi_{\rm cont}(x),
\label{Qexact_action_cont}
\ee
where
\be
\Xi_{\rm cont}(x)\equiv \tr\left[\frac14\eta [\phi,\,\bar{\phi}]  -i\sum_{\mu=1}^2 \psi_\mu D_\mu\bar{\phi}
+\chi (H-2i F_{12} )
\right].
\label{pre_action_cont}
\ee
We emphasize that the action (\ref{Qexact_action_cont}) 
is just a rewriting of (\ref{action_cont}) with a different notation,
and hence it is invariant under the full of the original supersymmetry transformations. 
The important point is that the $Q$-invariance is manifest in this form because 
$Q$ satisfies (\ref{Q_nilpotency_cont}) and $\Xi_{\rm cont}$ (\ref{pre_action_cont}) is gauge invariant.
It will be crucial for the lattice construction with keeping $Q$-symmetry discussed below.

\subsection{Lattice formulation with an exact supersymmetry}\label{subsec:unimproved lattice}
We briefly summarize the lattice formulation given in \cite{Sugino:2003yb}. 
The lattice action is defined in a $Q$-exact form, 
as (\ref{Qexact_action_cont}) for the continuum counterpart, 
and possesses the exact $Q$ supersymmetry invariance on the lattice.

Let us consider a two-dimensional square lattice 
with the periodic boundary conditions 
which is denoted by $\Lambda_L\equiv a\Z_L\times a\Z_L$, where
$a$ is the lattice spacing and $L$ is the number of lattice sites in each direction. 
The result of this paper is easily extended to the case of a rectangular lattice.
Hereafter, the lattice site is expressed as $x=(x_1,\,x_2)$, 
$x_\mu\in \{a,2a,\cdots, La\}$ ($\mu=1,2$). 
The fermions and scalars are defined on the sites, while 
the gauge fields are promoted to gauge group-valued variables $U_\mu(x)\in SU(N)$ defined on the link 
connecting $x$ and $x+a\hat{\mu}$ where $\hat \mu$ is the unit vector in the $\mu$-direction. 
For notational simplicity, we often use the same symbols for both of each continuum field and its lattice counterpart with keeping the mass dimensions. 
We put the subscripts ``cont'' and ``lat'' to distinguish them when needed.

The gauge transformations of the link fields $U_\mu$ and the other adjoint fields $F$ are given in the usual  manner:
\bea
&& U_\mu(x) \to \Lambda(x) \, U_\mu(x) \, \Lambda(x+a\hat\mu)^{-1}, 
\label{U_gaugetr}
\\
&& F(x) \to \Lambda(x) F(x) \Lambda(x)^{-1},
\label{F_gaugetr}
\eea
where $\Lambda(x)=e^{i\omega(x)} \in SU(N)$ is a gauge transformation function. 
For later use, we introduce 
the gauge covariant forward (backward) difference operator $\nabla_\mu$ ($\nabla_\mu^*$) as  
\bea
&& \nabla_\mu F(x) \equiv \frac{1}{a}\left(U_\mu(x)F(x+a\hat{\mu}) U_\mu(x)^{-1}-F(x)\right),
\label{forward_diff}\\
&& \nabla^*_\mu F(x) \equiv \frac{1}{a}\left(F(x) - U_\mu(x-a\hat{\mu})^{-1}F(x-a\hat{\mu}) U_\mu(x-a\hat{\mu})\right).
\label{backward_diff}
\eea
The covariant difference operators (\ref{forward_diff}) and 
(\ref{backward_diff}) are covariant under the lattice gauge transformations,
(\ref{U_gaugetr}) and (\ref{F_gaugetr}), 
as their names suggest. The plaquette field,
\begin{equation}
U_{\mu\nu}(x)=U_\mu(x)U_\nu(x+a\hat{\mu})U_\mu(x+a\hat{\nu})^{-1} U_\nu(x)^{-1},
\label{plaquette}
\end{equation}
is another important gauge covariant quantity.

If we use a naive relation
$U_\mu(x) =e^{iaA_\mu(x)}$
and send $a$ to zero,  
\footnote{More detailed analysis with an improved relation 
(\ref{U_mid-point}) is given in the next subsection. } 
the difference operators
(\ref{forward_diff}) and (\ref{backward_diff}) coincide with the correct covariant derivative 
(\ref{covariant_derivative}), and the plaquette field reproduces
the continuum field tensor (\ref{field_tensor}) as
$U_{12}(x) =1 +i a^2F_{12}(x) +\cO(a^3)$.
Moreover, the infinitesimal form of the lattice gauge transformations,
\bea
\delta_\omega U_\mu(x) = -ia\nabla_\mu \omega(x) U_\mu(x), 
\qquad \delta_\omega F(x) =i [\omega(x), F(x)], 
\label{infinitesimal_lattice_gauge_transformation}
\eea
also reproduce the correct continuum limit (\ref{gauge_transformation}).

The $Q$-transformation is realized on the lattice  \cite{Sugino:2003yb}  by 
\bea
& & QU_\mu(x) = ia\psi_\mu(x)\,U_\mu(x), \label{naive_QU} \\
& & Q\psi_\mu(x) = i\nabla_\mu\phi(x)+ ia\psi_\mu(x)\psi_\mu(x),  \label{naive_Qpsi} \\
& & Q\phi(x) = 0,  \label{naive_Qphi} \\
& & Q\bar{\phi}(x) = \eta(x), \qquad Q\eta(x)=[\phi(x), \,\bar{\phi}(x)], \label{naive_Qeta} \\
& & Q\chi(x) = H(x), \qquad QH(x) = [\phi(x),\,\chi(x)]. \label{naive_Qchi}
\eea
Note that the transformation above
remains nilpotent up to an infinitesimal lattice gauge transformation
(\ref{infinitesimal_lattice_gauge_transformation})
 with the parameter $\phi(x)$ 
on the lattice:
\be
Q^2 =-i\delta_\phi. 
\label{Q_nilpotency_lat}
\ee
It reproduces the transformation rule in the continuum theory, 
(\ref{cont_QA})-(\ref{cont_Qeta}), after taking the continuum limit. 
With use of (\ref{naive_QU})-(\ref{naive_Qchi}), the action (\ref{Qexact_action_cont}) is transcribed to the lattice action: 
\begin{align}
S_{\rm lat} \equiv Q\,\frac{a^2}{2g^2} \sum_{x\in \Lambda_L} \, \Xi_{\rm lat}(x)  
\label{S_lat}
\end{align}
with 
\begin{align}
\Xi_{\rm lat}(x) \equiv \tr \biggl\{& \frac{1}{4} \eta(x) [\phi(x),\bar\phi(x)]
 -i \sum_{\mu=1}^2 \psi_\mu(x) \nabla_\mu \bar\phi(x) 
 +\chi(x) \left( H(x) - \frac{i}{a^2} \Phi_{\rm TL}(x) \right) \biggr\},
 \label{pre action lat}
\end{align}
where $\Phi_{\rm TL}(x)$ is a lattice version of the field tensor (\ref{field_tensor}) satisfying 
$\Phi_{\rm TL}(x) \rightarrow 2a^2 F_{12}(x)$ as $a \rightarrow 0$, and explicitly given in what follows.   

In discussing the continuum limit, the plaquette field is usually expanded around unity. 
In the current case, we need to be careful about this point. 
After integrating the auxiliary field $H$, 
the lattice  gauge action becomes
\be
S_G =\frac{1}{8g^2a^2} \sum_{x \in \Lambda_L} \tr(\Phi_{\rm TL}^2(x) ), 
\label{gauge_action}
\ee
and one finds that
\be
\Phi_{\rm TL}(x) =0, \quad {\rm for \ all} \  x \in \Lambda_L,
\label{vacuum_eq}
\ee
gives the vacuum configurations corresponding to the continuum limit  ($a$ goes to zero with $g$ fixed). 
If the vacuum configuration is unique and gives $U_{12}(x)={\bf 1}_N$, one can expand the plaquette field around unity 
and confirm that the lattice action reproduces the continuum one.
If it is not the case, the lattice action is not guaranteed to provide the desired continuum action in general.
For instance, for $SU(2)$, a naive choice
$
\Phi_{\rm TL}(x) = - i \left(U_{12}(x)-U_{21}(x) \right)
$
does not lead to the unique vacuum  $U_{12}(x)={\bf 1}_2$. 
Actually,  another vacuum $U_{12}(x)=-{\bf 1}_2$ satisfies (\ref{vacuum_eq}) as well 
and causes an obstacle for taking the correct continuum limit. 
In order to reproduce the correct continuum gauge action, 
$\Phi_{\rm TL}(x)$ should be chosen so that the gauge action (\ref{gauge_action}) has no nontrivial minima 
other than $U_{12}(x)={\bf 1}_N$.  

Two possibilities for desired lattice field tensor have been discussed in~\cite{Sugino:2004qd} and \cite{Matsuura:2014pua}. 
Since the traceless field $\Phi_{\rm TL}$ is generally given by
\be
\Phi_{\rm TL}(x) \equiv \Phi(x) - \left(\frac{1}{N}\tr\Phi(x)\right){\bf 1}_N,
\ee
we can use $\Phi$ for defining $\Phi_{\rm TL}$ .
One possibility is 
\bea
\Phi_1(x) =
\left\{
\begin{array}{ll}
\frac{-i(U_{12}(x)-U_{21}(x))}{1-\frac{1}{\epsilon^2}||1-U_{12}(x)||^2} 
&  {\rm for} \ \  ||1-U_{12}(x)||<\epsilon,
\\
\infty  &   {\rm otherwise},
\end{array}
\right.
\label{Phi_admissibility}
\eea
where $\epsilon$ is a positive number chosen in the range 
$0<\epsilon <2\sqrt{2}$ for $N=2,3,4$ and $0<\epsilon <2\sqrt{N}\sin(\frac{\pi}{N})$ for $N\geq 5$. The other possibility is   
 \bea
 \Phi_2(x) = 
 \frac{4i}{M}\cdot 
 \frac{2 - U_{12}(x)^M - U_{21}(x)^{M}}{U_{12}(x)^M - U_{21}(x)^{M} },
\label{Phi_tanM} 
 \eea
with $M$ being an integer satisfying $2M\ge N$. 
When $M$ is even ($M=2m$), (\ref{Phi_tanM}) can be recast as 
\bea
\Phi_2(x) =  
-\frac{2i}{m} \cdot 
\frac{U_{12}(x)^m - U_{21}(x)^m}{ U_{12}(x)^m + U_{21}(x)^m},
 \label{Phi_tan2m}
\eea 
which would be more convenient for numerical simulation. 

Note that the r.h.s. in  (\ref{Phi_tanM}) is a hermitian matrix since the denominator and the numerator commute with each other.
In both cases, only the single vacuum $U_{12}={\bf 1}_N$ is allowed, 
and by expanding the plaquette field around unity, we can verify that $\Phi_{i,\rm TL}(x)=2a^2F_{12}(x) +\cO(a^3)$.

Thus, we can construct the continuum limit of the lattice action (\ref{S_lat}) with (\ref{Phi_admissibility}) or (\ref{Phi_tanM}) 
and verify that it does coincide with the action (\ref{action_cont}) via 
 (\ref{Qexact_action_cont}) and (\ref{pre_action_cont}). 
The lattice action is exactly invariant under the lattice $Q$-transformation (\ref{naive_QU})-(\ref{naive_Qchi})
 thanks to the nilpotency of $Q$, (\ref{Q_nilpotency_lat}).  The perturbative power counting theorem tells us that any relevant supersymmetry breaking operators are forbidden by the $Q$-symmetry and the internal $U(1)_R$-symmetry\cite{Sugino:2003yb}, and all the supersymmetries are shown to be restored at least in the perturbation theory.  

The lattice action can be used for the numerical simulations, 
since the same forward difference operators employed in (\ref{naive_Qpsi}) and  (\ref{pre action lat}) yield a semi-positive 
boson action which is suitable for the Monte-Carlo method. Numerical results indicate that the restoration of the full supersymmetries 
does occur beyond the perturbation theory \cite{Kanamori:2008bk,Kanamori:2008yy,Hanada:2009hq,Hanada:2010qg}.

\subsection{Classical continuum limit}
\label{sec:naive_limit_of_original_action}
The action 
(\ref{action_cont}) is correctly reproduced from  the lattice action (\ref{S_lat}) as we take the lattice spacing to zero,
as seen in the previous subsection. 
In this subsection, in order to find a proper ${\cal O}(a)$ improvement procedure,  
we study the classical continuum limit more precisely 
by expanding various quantities (difference operators, plaquette field, and $Q$-transformation) 
in the lattice spacing,
and determine the order at which the first deviation terms 
from the continuum theory appear in the expansion.

The lattice fields are not defined on the continuum spacetime but on the lattice. 
To expand the fields in the lattice spacing $a$, we first embed the lattice in the continuum spacetime 
and regard the lattice functions as smooth functions defined on the continuum 
spacetime.
In order to associate the link fields to the continuum gauge fields, we employ the midpoint prescription 
\begin{equation}
U_\mu(x)= e^{iaA_\mu(x_c)}, 
\label{U_mid-point}
\end{equation}
where 
$
x_c=x + (a/2) \hat \mu
$. 
The link field can be interpreted as the Wilson line, 
\begin{align}
&U_\mu(x) ={\rm P} \,   e^{i \int_0^{a} dt \,A_\mu(x+t\hat{\mu}) } \nn \\
& \hspace{10mm}  \equiv  1 + i\int_0^a dt\,A_\mu(x+t\hat{\mu}) + i^2\int_0^a dt \int_0^a  ds\,\theta(s-t)\,A_\mu(x+t\hat{\mu})A_\mu(x+s\hat{\mu})+\cdots,
\label{U_pathorder}
\end{align}
when the lattice is embedded in the continuum spacetime. 
(\ref{U_mid-point}) is obtained by expanding the r.h.s. in $a$ up to $\cO(a^2)$.
Namely, the midpoint prescription reproduces the continuum gauge transformation up to this order, 
$U_\mu(x)= e^{iaA_\mu(x_c)+\cO(a^3)}$.
It is straightforward to increase the precision by proceeding the expansion in $a$.~\footnote{ 
For example, we have 
\[
U_\mu(x) = \exp\left[iaA_\mu(x_c) +i\frac{a^3}{24}\partial_\mu^2A_\mu(x_c) 
-\frac{a^3}{12}[A_\mu(x_c),\,\partial_\mu A_\mu(x_c)] +\cO(a^4)\right]. 
\]
} 
We identify the other adjoint lattice fields (scalar and fermi fields) as the fields on the embedded points where they are defined.
Under these identifications, we obtain smooth functions associated with the lattice fields
by smoothly interpolating those fields on entire continuum spacetime.  
Thus, the expansion with respect to the lattice spacing can be simply performed by the Taylor expansion.
For instance, we can expand a function $f(x)$ around the midpoint $x_c=x + (a/2) \hat \mu$:
\begin{eqnarray}
f(x) = f(x_c) - \frac{a}{2} \partial_\mu f(x_c)   +  \frac{a^2}{8} \partial^2_\mu f(x_c)   + \cO(a^3).
\label{identity_expansion}
\end{eqnarray}
By using the identities for $\nabla_\mu$ and  arbitrary deformation $\delta$:
\be
\nabla_\mu f(x) = D^c_{\mu} \, f(x_c) +\frac{ia}{2}[A_\mu(x_c), D^c_\mu\, f(x_c)] + \cO(a^2) 
\label{identity_nabla_f}
\ee
with $D_\mu^c=\partial_\mu + i[A_\mu(x_c), \, \cdot\, ]$, 
and 
\be
\delta A = \delta e^A \cdot e^{-A} -\frac{1}{2} [A,  \delta e^A \cdot e^{-A}] + \cO(A^3),
\label{identity_delta_G}
\ee
we find that the lattice gauge transformation (\ref{infinitesimal_lattice_gauge_transformation}) is automatically $\cO(a)$-improved as 
\begin{eqnarray}
\delta_{{\rm lat}, \,\omega} = \delta_{{\rm cont},\,\omega} + \cO(a^2).
\label{gauge_order_a}
\end{eqnarray}

On the other hand, the $\cO(a)$ improvement of the lattice $Q$-transformation  
(\ref{naive_QU})-(\ref{naive_Qchi}) is not automatic.
Indeed, 
the $Q$-transformation of $\psi_\mu$  (\ref{naive_Qpsi}) has a quadratic $\cO(a)$-term of the fermion 
and the unimproved forward difference operator that is expanded as
\begin{eqnarray}
\nabla_\mu  = D_\mu  +\frac{a}{2} D_\mu^2  + \cO(a^2).
\label{fwd_order_a}
\end{eqnarray} 
The $Q$-transformation of the gauge fields
also has the non-zero $\cO(a)$-correction.
Hence, the total $Q$-transformation satisfies
\begin{eqnarray}
Q_{\rm lat}= Q_{\rm cont} + \cO(a),
\label{dev Q naive}
\end{eqnarray}
although those of  the other fields retain their continuum forms.  
\footnote{
Somewhat surprisingly, the $\cO(a)$-term in (\ref{dev Q naive})  vanishes when applying the lattice $Q$-transformation twice, 
because both the continuum and the lattice gauge transformations with the parameter $\phi$ satisfy (\ref{gauge_order_a}). }

Similarly, the integrand $\Xi_{\rm lat}$ (\ref{pre action lat}) behaves as
 \begin{equation}
  \Xi_{\rm lat}(x)= \Xi_{\rm cont}(x) + {\cal O}(a) , 
  \label{dev Xi naive}
\end{equation}
since the $\cO(a)$-terms come from the forward difference operator (\ref{fwd_order_a}) and the unimproved plaquette field, 
\begin{align}
U_{12}(x)=\exp\left( 
 ia^2 F_{12}(x) 
 +i\frac{a^3}{2}(D_1 + D_2) F_{12}(x) 
+{\cal O}(a^4)
\right).
\label{expansion_naive_plaquette}
\end{align}
Note that the $\cO(a)$-correction does not cancel even if we take 
the combination $U_{12}-U_{21}$ as in $\Xi$. 

We thus find that the total lattice action is
\begin{eqnarray}
S_{\rm lat}=S_{\rm cont} + {\cO}(a),
\end{eqnarray}
because of (\ref{dev Q naive}) and (\ref{dev Xi naive}) with (\ref{S_lat}).  
In order that the lattice action coincides with the continuum action up to $\cO(a^2)$ terms,
we have to improve 
the covariant difference operator, the field tensor
and the lattice $Q$-transformation
with keeping the nilpotency.

\section{Method of tree-level ${\cO}(a)$ improvement}\label{sec:improvement}
In the last section, we have seen that 
the lattice gauge transformation is already $\cO(a)$-improved 
while $Q$ and $\Xi_{\rm lat}$ have the $\cO(a)$-corrections, (\ref{dev Q naive}) and  (\ref{dev Xi naive}), 
and thus the resultant action  (\ref{S_lat}) produces the continuum one up to $\cO(a)$ terms.
In this section, we will explain our strategy to improve $Q$ and 
$\Xi_{\rm lat}$ so that 
$S^{imp}_{\rm lat}=S_{\rm cont} + {\cO}(a^2)$
is obtained. 
The explicit form of the tree-level ${\cal O}(a)$-improved action will be given in the next section.

\subsection{Locality}\label{subsec:locality}
Before going to the detail of the strategy, we should mention on 
the locality condition we employ throughout this paper. 
Indeed, 
the improvement will be performed with keeping the locality 
of the $Q$-transformation in the sense that 
long-range interactions in the lattice unit are suppressed, 
at least, by the exponential of the distance.

First, let us define ultra-local operators and exponentially local ones. 
Suppose that the lattice size is infinite and the sites are labeled by integers, $x_\mu \in a {\mathbb Z}$, and 
$\varphi_\alpha(x)$ is a lattice field with finite internal index $\alpha=1,2,\cdots, N_{i}$.
In the present case, $N_i=N^2-1$ since $\varphi(x)$ is expanded by a basis of $su(N)$ generators: 
$\varphi(x)=\sum_{\alpha=1}^{N^2-1}\varphi_\alpha(x)T_\alpha$.
An operator $R$ acting to $\varphi(x)$ is  formally represented as a kernel:
\be
(R\varphi)_\alpha(x) = \sum_{y,\,\beta}R_{\alpha\beta}(x,y)\varphi_\beta(y). 
\label{eq:kernel_R}
\ee
The kernel is an infinite dimensional matrix with the row $\alpha,\,x$ and the column $\beta,\,y$. 
In the following and in appendix~\ref{app:R}, 
$R(x,y)$ denotes the $N_i\times N_i$ matrix with respect to the internal index.

The ultra-local operator is defined by
\be
R(x,y)=0, \quad {\rm for} \ \ ||x-y||_1 > ar,
\label{condition_ultra_local}  
\ee
where the localization range $r$ is 
a fixed natural number.~\footnote{
As a definition of the distance, we use the ``taxi driver distance'': 
$||x-y||_1\equiv \sum_{\mu}|x_\mu-y_\mu|$. For the Euclid distance 
$||x-y||_2\equiv\sqrt{\sum_\mu(x_\mu-y_\mu)^2}$, the following argument will also be similar. 
}
For instance, the forward and the backward difference operators are ultra-local with the localization range one.
The kernel of an ultra-local operator forms a banded diagonal matrix, which is suited for numerical applications.  

On the other hand, $R$ is referred to as an exponentially local operator 
if there exist positive constants $C$ and $\kappa$ such that 
\be
|R_{\alpha\beta}(x,y)|\leq C \,e^{-\frac{\kappa||x-y||_1}{a}}.
\label{condition_exp_local}  
\ee
As is well-known, the overlap Dirac operator~\cite{Neuberger:1997fp,Neuberger:1998wv} satisfies this type of locality~\cite{Hernandez:1998et}. 
It is obvious that any ultra-local operator satisfies (\ref{condition_exp_local}), but the converse is not true in general. 
The exponential locality is therefore a milder condition as the locality. 

In the continuum limit, both locality conditions reproduce a local continuum field theory with finite number of derivatives 
contained in its classical action, 
because there are no 
contributions from $y$ being separated from $x$ by a finite physical length as $a \rightarrow 0$. 
Also in the point of view of the renormalization group and the universality hypothesis~\cite{Wilson:1973jj,Niedermayer:1998bi,Hernandez:1998et}, 
the exponential locality is allowed as a locality condition of the lattice theory having the desired continuum limit.  
We therefore employ the exponential locality to construct the $\cO(a)$-improved theory in this paper.

\subsection{$Q$-transformation}
\label{subsec:SUSY improve}

Among the $Q$-transformations \eqref{naive_QU}--\eqref{naive_Qchi},
we have to improve only the two transformations $QU_\mu$ (\ref{naive_QU}) and $Q\psi_\mu$ (\ref{naive_Qpsi}) 
since the others (\ref{naive_Qphi})-(\ref{naive_Qchi})
coincide with their continuum forms and are irrelevant to the present purpose.

Here we should point out that, 
even if we give an improved transformation of $U_\mu$, 
it is not clear {\it a-priori} 
if we can define a modified transformation of $\psi_\mu$ that is 
consistent with the nilpotency of $Q$ (\ref{Q_nilpotency_lat}).
Furthermore, even though it is possible, it is not yet clear whether the transformation 
of $\psi_\mu$ satisfies the locality. 
The following lemma is crucial to give answers to these points: 
\begin{lemma}
Let $f_\mu(x)$ 
be a function that depends on $(U_\lambda,\psi_\lambda,\phi)$ with the same gauge transformation property as $U_\mu(x)$,  and suppose 
that
\be
QU_\mu(x)= f_\mu(x). 
\label{lemma_QU}
\ee
Let $g_\mu(x)$  
be a function that depends on $(U_\lambda,QU_\lambda,\phi)$  with the same gauge transformation property as $\psi_\mu(x)$. 
If (\ref{lemma_QU}) can be solved in term of $\psi_\mu(x)$ as follows: 
\be
\psi_\mu(x) = g_\mu (x),
\label{lemma_psi}
\ee 
then one can consistently define the $Q$-transformation by (\ref{lemma_QU}) and 
\bea
Q\psi_\mu(x) & = & \sum_{y,\,\nu}\left[QU_\nu(y)\cdot \frac{\delta}{\delta U_\nu(y)} -i\delta_\phi U_\nu(y)\cdot \frac{\delta}{\delta(QU_\nu(y))}\right]
g_\mu (x),
\label{lemma_Qpsi} \\
Q\phi(x) & = & 0,
\eea 
so that $Q^2=-i\delta_\phi$ for all fields.
\end{lemma}
Note that, for the original transformations (\ref{naive_QU}) and (\ref{naive_Qpsi}),
the r.h.s. of $QU_\mu(x)$ actually defines such a function $f_\mu(x)$.
In that case, $g_\mu(x) \equiv -\frac{i}{a}(QU_\mu(x))U_\mu(x)^{-1}$ that is identical to $\psi_\mu(x)$. 
This lemma actually holds in general frameworks with the exact gauge invariance, including but not limited to the lattice gauge theory.
We give a proof in appendix~\ref{app:lemma} and simply use the result here. 
\footnote{
The essential point of the proof is that the r.h.s. of the expression (\ref{lemma_Qpsi}) 
is  identified with $Q g_\mu$ once $-i\delta_\phi U_\nu$ in the second 
term is replaced by $Q^2 U_\nu$. This implies $Q^2=-i \delta_\phi$ for $U_\mu$.
The nilpotency is also satisfied for  $QU_\mu$
since $Q$ commutes with $\delta_\phi$, that is,
 $Q^2(QU_\mu)=Q(Q^2U_\mu)=-iQ\delta_\phi U_\mu=-i\delta_\phi (QU_\mu)$.
In addition, we find that $Q^2 = -i \delta_\phi$ for $\psi_\mu$ because $\psi_\mu$ is given by $g_\mu$ that is a function of $U_\nu$ and $QU_\nu$ 
for which $Q^2=-i\delta_\phi$ was already satisfied. 
}

Let us consider a possible deformation of $Q$ based on this lemma. 
Suppose that an improved transformation takes the form, 
\begin{eqnarray}
Q^{imp} U_\mu(x) = ia (\cR_\mu \psi_\mu(x)) U_\mu(x),
\label{general_QU}
\end{eqnarray}
where ${\cal R}_\mu$ is an operator which improves the $Q$-transformation of $U_\mu(x)$. 
In this case, $g_\mu$ in the lemma is formally given by
\begin{eqnarray}
g_\mu= \cR_\mu^{-1} \left\{ -\frac{i}{a}(Q^{imp}U_\mu) U_\mu^{-1}\right\}, 
\label{g_mu_general}
\end{eqnarray}
to express $\psi_\mu$ as (\ref{lemma_psi}) and we have the correct transformation (\ref{lemma_Qpsi}). 
The transformation of $\psi_\mu$ thus becomes 
\begin{eqnarray} 
Q^{imp}\psi_\mu(x) = i\cR_\mu^{-1} \nabla_\mu \phi(x) + ia \sum_{y,\nu}  (\cR_\nu \psi_\nu(y)) U_\nu(y)\cdot \frac{\delta }{\delta U_\nu(y)}g_\mu(U_\lambda,Q^{imp}U_\lambda;x).
\label{general_Qpsi}
\end{eqnarray} 
Here we find that one cannot use an arbitrary $\cR_\mu$, because 
$\cR_\mu$ must not have the zero-eigenvalue in order to have a well-defined inverse.

Keeping these points in mind, let us consider the improvement of 
$Q U_\mu$ and $Q \psi_\mu$ in detail. 
In contrast to the case of the gauge transformation (\ref{gauge_order_a}),
the reason why the original transformation 
(\ref{naive_QU}) has an $\cO(a)$ correction 
is simple:  
The gauge field $A_\mu(x)$ is defined on the midpoint
while the corresponding fermionic field $\psi_\mu(x)$ is defined on a site. 
Hence a heuristic choice of the improved transformation would be
\begin{eqnarray}
Q^{imp}_{naive} U_\mu(x) = \frac{ia}{2} \left(\psi_\mu(x)\,U_\mu(x) + U_\mu(x) \psi_\mu(x+a\hat\mu) \right), 
\label{heuristic_choice}
\end{eqnarray}
which corresponds to 
\be
\cR_\mu^{\rm naive} = 1 + \frac{a}{2} \nabla_\mu. 
\label{naive_Rmu}
\ee
Indeed, we can show that (\ref{heuristic_choice})  turns out to have the desirable property 
\begin{eqnarray}
Q^{imp}_{naive} A_\mu(x) = \psi_\mu(x) + \cO(a^2)
\label{necessary cond}
\end{eqnarray}
from (\ref{identity_expansion}), (\ref{identity_nabla_f}) and (\ref{identity_delta_G}) with the midpoint prescription (\ref{U_mid-point}). 
However, this choice is too naive. 
In fact, since the eigenvalues of $a\nabla_\mu$ are parametrized by 
\be
-1+e^{i\theta} \qquad (\theta\in \mathbb{R}), 
\label{nabla_EV}
\ee
(\ref{naive_Rmu}) has zeros in its spectrum.  
\footnote{
(\ref{nabla_EV}) is understood from the identity $a^2\nabla_\mu^*\nabla_\mu=a\nabla_\mu-a\nabla_\mu^*$ and 
the relation $\nabla_\mu^*=-\nabla_\mu^\dagger$ with respect to the inner product of $su(N)$-valued functions: 
$(f, g)\equiv \sum_{x\in\Lambda_L}\tr\left(f(x)^\dagger\,g(x)\right)$.
$\nabla_\mu$ and $\nabla_\mu^*$ mutually commute and can be simultaneously diagonalized by a unitary transformation.}

This suggests that
\begin{eqnarray}
\cR_\mu = 1+ \frac{a}{2} \nabla_\mu -r a^2 \nabla_\mu \nabla_\mu^*   \quad (r > 0)
\label{R_solution}
\end{eqnarray} 
is a candidate for our desired solution improving $Q$-transformations up to $\cO(a)$. 
In contrast to (\ref{naive_Rmu}), the third term (the Wilson-term) 
lifts the zero-mode with keeping the small $a$ behavior unchanged 
as long as $r\neq 0$. 
(Note that the case $r<0$ is possible for the purpose of this section, but it 
will be rejected because it leads to an incorrect fermion measure as we will see in section \ref{sec:final result}.)
In this case, $\cR^{-1}_\mu$ is exponentially local, because one can show that an ultra-local operator $\cR_\mu$, 
whose spectrum is bounded by an upper and a non-zero lower bounds, has an inverse being exponentially local. 
Details including the definitions of the terminologies are presented in appendix~\ref{app:locality}.~\footnote{ 
This situation reminds us the locality of the overlap Dirac operator. Unlike the Wilson operator $D$ that is ultra-local, the overlap operator is given by the inverse square root of $D^\dag D$ of which the locality is not immediately obvious.    
In \cite{Hernandez:1998et}, it is proven that the overlap operator is indeed local with exponentially decaying tails if the eigenvalues of $D^\dag D$ are bounded from above and below.  
We can apply the same logic to the present case. }
We thus have the $\cO(a)$-improved transformation for the link fields within the locality principle.

The lemma also guarantees that 
the $Q$-transformation of the fermi fields $\psi_\mu$ 
\eqref{lemma_Qpsi} 
are automatically improved 
in the case that both of the $Q$-transformation and 
the gauge transformation $\delta_\phi$ of $U_\mu$ are improved. 
In fact, (\ref{lemma_QU}) and (\ref{lemma_psi}) have the same information, 
and thus the expansion of $g_\mu$ in $a$ should reproduce the counterpart of the continuum theory 
without the error of $\cO(a)$, from the assumption on the improvement of $U_\mu$.

Of course, we can explicitly verify that (\ref{general_Qpsi}) coincides to that of the continuum theory (\ref{cont_Qpsi}) 
up to $\cO(a)$.  
In the improved transformation, the forward difference operator in the original transformation (\ref{naive_Qpsi}) 
is replaced by $\cR_\mu^{-1} \nabla_\mu $ that provides the improvement because of the expansion 
\begin{eqnarray}
\cR_\mu^{-1} \nabla_\mu = D_\mu + \cO(a^2), 
\label{RandD}
\end{eqnarray}
and the fermion bilinear term in (\ref{naive_Qpsi}) does not appear at the order of $a$ as is seen in the following. 
We may use $1 + \frac{a}{2} \nabla_\mu$ and $1 - \frac{a}{2} \nabla^*_\mu$ as 
$\cR_\mu$ and $\cR_\mu^{-1}$ respectively, within the precision of $\cO(a)$. 
Then, the second term of (\ref{general_Qpsi}) is shown to be
\begin{eqnarray}
\frac{i}{2a} \left\{  Q^{imp}U_\mu(x) Q^{imp}U^{-1}_\mu(x)  -Q^{imp}U^{-1}_\mu(x-a\hat\mu)  Q^{imp}U_\mu(x-a\hat\mu) \right\} + \cO(a^2).
\end{eqnarray}
Although this seems to be $\cO(a)$ at the first sight since $QU_\mu$ is of the order the lattice spacing, 
the leading contributions in the parenthesis cancel each other. 
We thus see
that (\ref{general_Qpsi}) is improved up to the desired order.

Therefore, for given $\cR_\mu$, (\ref{R_solution}) or any other choices listed in appendix~\ref{app:R},  
the transformations (\ref{general_QU}) and  (\ref{general_Qpsi}) with (\ref{naive_Qphi})-(\ref{naive_Qchi}) are local and 
satisfy both (\ref{Q_nilpotency_lat}) and
\begin{eqnarray}
Q^{imp}_{\rm lat} = Q_{\rm cont} +\cO(a^2), 
\end{eqnarray}
for all the fields.

\subsection{The integrand $\Xi$}\label{subsec:differece improve}
In this section, we will improve  $\Xi_{\rm lat}$, in particular, the difference operator and the lattice field tensor 
appearing there.

As we have seen in (\ref{RandD}), the forward difference operator (\ref{fwd_order_a}) is improved 
by multiplying $\cR_\mu^{-1}$. 
Since the plaquette field is similarly expanded as (\ref{expansion_naive_plaquette}), 
its improvement is achieved analogously:
\be
U_{\mu\nu}^{imp}(x) \equiv \cR_{12}U_{\mu\nu}(x),
\label{U12_imp}
\ee
where we adopt  
 \be
\cR_{12}=1-\frac{a}{2}(\nabla_1+\nabla_2).
\label{R12_lat}
\ee
Note that the improvement factor $\cR_{12}$ is common for $U_{12}$ and $U_{21}$, 
but \eqref{R12_lat} is not the unique choice.
Other possible choices of $\cR_{12}$ are given in appendix~\ref{app:R}.

The improved $\Xi$ is thus given by
\begin{equation}
\Xi_{\rm lat}^{imp}
= \tr \biggl\{ \frac{1}{4} \eta [\phi,\bar\phi]
 -i \sum_{\mu=1}^2 \psi_\mu \,\nabla_\mu^{imp} \bar\phi
+\chi \left( H - \frac{i}{a^2} \Phi_{\rm TL}^{imp} \right) \biggr\},  
 \label{Xi_lat_imp_org}
\end{equation}
where 
\begin{eqnarray}
&&  \nabla_\mu^{imp} = \cR_\mu^{-1} \nabla_\mu, 
\label{imp_nabla_in_xi}
\\
&&\Phi^{imp}_{\rm TL}(x) = \cR_{12}\Phi_{\rm TL}(x),
\end{eqnarray}
for the unimproved $\Phi_{\rm TL}$, (\ref{Phi_admissibility}) or (\ref{Phi_tanM}). 
Here, $\cR_\mu$ appearing in (\ref{imp_nabla_in_xi}) should be taken as the same as those used in the $Q$-transformations, (\ref{general_QU}) and (\ref{general_Qpsi}),
in order to make the bosonic part of the action semi-positive.
Note that, since $\cR_{12}$ is invertible, 
the field equation $\Phi_\textrm{TL}^{imp}=0$ is identical with (\ref{vacuum_eq}).
Therefore we can use the same choice (\ref{Phi_admissibility}) or (\ref{Phi_tanM}) to forbid the extra vacua even in this case.

Now it is easy to see that 
\begin{align}
  \Xi^{imp}_{\rm lat}(x) &= \Xi_{\rm cont}(x) + {\cal O}(a^2).
  \label{dev Xi imp}
 \end{align}
The improved theory is defined by both the improved $Q$-transformation (\ref{general_QU}) and  (\ref{general_Qpsi}) with 
 (\ref{R_solution})
and the improved $\Xi(x)$  (\ref{Xi_lat_imp_org}).

\section{Fermion doublers and path-integral measures in improved theory}
 \label{sec:final result}
In this section, we first summarize the ${\cal O}(a)$-improved lattice action 
explained in the previous section.
After that, we show the absence of the fermion doublers 
and give appropriate $Q^{imp}$-invariant path-integral measures.

\subsection{Summary of the improved theory}
\label{subsec:result}

The improved action obtained in the previous section is given by
\bea
S^{imp}_{\rm lat} & =&  Q^{imp} \frac{a^2}{2g^2} \sum_{x\in\Lambda_{L}} \Xi^{imp}_{\rm lat}(x) , 
\label{improved action}
\\
\Xi_{\rm lat}^{imp}
& =  & \tr \biggl\{ \frac{1}{4} \eta [\phi,\bar\phi]
 -i  \sum_{\mu=1}^2 \psi_\mu \,\cR_\mu^{-1} \nabla_\mu \bar\phi
+\chi \left( H - \frac{i}{a^2} \cR_{12}\Phi_{\rm TL} \right) \biggr\},
 \label{Xi_lat_imp}
\eea
where $\cR_\mu$ and $\cR_{12}$ are given, for example, by 
\begin{eqnarray}
&& \cR_\mu = 1+ \frac{a}{2} \nabla_\mu -r a^2 \nabla_\mu \nabla_\mu^*   \quad (r > 0),
\label{R_mu_final}\\
&& \cR_{12}=1-\frac{a}{2}(\nabla_1+\nabla_2).
\label{R12_final}
\end{eqnarray}
Note that, 
as mentioned in the previous section, 
these factors are not unique and other possible choices 
are summarized in appendices \ref{app:Rmu} and \ref{app:R12}.

The improved $Q$-transformations are expressed as 
\begin{eqnarray}
&&Q^{imp} U_\mu(x) = ia \psi'_\mu(x) U_\mu(x),
\label{QU_final}\\
&&
Q^{imp} \psi'_\mu(x) = i\nabla_\mu\phi(x) + ia \psi'_\mu(x) \psi'_\mu(x),  
\label{Qpsi_final}\\
&& Q^{imp} \phi(x) = 0,  \label{Qphi_final} \\
&& Q^{imp} \bar{\phi}(x) = \eta(x), \qquad Q^{imp} \eta(x)=[\phi(x), \,\bar{\phi}(x)], \label{Qeta_final}\\
&& Q^{imp} \chi(x) = H(x), \qquad Q^{imp} H(x) = [\phi(x),\,\chi(x)] \label{Qchi_final}
\end{eqnarray}
in terms of 
\be
\psi'_\mu(x) \equiv \cR_\mu\psi_\mu(x).
\label{psi-prime}
\ee
The first two transformations are built in section \ref{subsec:SUSY improve}, 
(\ref{general_QU}) and  (\ref{general_Qpsi}) with (\ref{g_mu_general}) for 
a given $\cR_\mu$.  
The others are the same with the continuum transformations 
(\ref{naive_Qphi})-(\ref{naive_Qchi}). 
These transformations obey $(Q^{imp})^2=-i\delta_\phi$  
as in the unimproved theory, and the improved action retains the invariance of the exact supersymmetry.%
~\footnote{
Conversely, once $Q^{imp} U_\mu$ is defined by (\ref{QU_final}), the transformation law for $\psi_\mu$ is uniquely determined under the constraint  $(Q^{imp})^2=-i\delta_\phi$. } 
Note that the improved transformations are identical with the original transformations 
(\ref{naive_QU})-(\ref{naive_Qchi}) under the replacement of $\psi_\mu$ by $\psi'_\mu$. 

The improved action (\ref{improved action}) coincides with the unimproved lattice action given in \cite{Sugino:2003yb} 
for  $\cR_\mu=\cR_{12}=1$, 
and the $\cO(a)$ terms of $\cR_\mu$ and $\cR_{12}$ provide the improvement. 
As we have already shown in the last section, $Q^{imp}$ and $\Xi^{imp}$ reproduce the continuum $Q$ and $\Xi$ 
with no $\cO(a)$ error, respectively. So, the improved action indeed satisfies $S^{imp}_{\rm lat} = S_{\rm cont} +\cO(a^2)$ 
as $a \rightarrow 0$.

As with the unimproved theory, the action (\ref{improved action}) is exactly invariant under the axial $U(1)_R$-transformation~\cite{Sugino:2003yb} (as well as $Q^{imp}$), 
because the multiplication by $\cR_\mu$ or $\cR_{12}$ does not affect the transformation properties.  
So we can conclude that the same perturbative arguments hold in this case, 
and all of the supersymmetries are restored in the continuum limit at the quantum level, at least, in the perturbation theory.

In order to define the quantum theory, 
we need to specify not only the action but also the path-integral measure.  
There are two candidates; 
the natural measure, 
\begin{equation}
	D\varphi_\textrm{natural} \equiv \prod_{x} d\phi(x)  d\bar\phi(x) dH(x) \, 
d\chi(x) d\eta(x) \prod_\mu dU_\mu(x)d\psi_\mu(x), 
\label{naive measure}
\end{equation}
and the manifestly $Q^{imp}$-invariant measure, 
\bea
D\varphi & \equiv & 
 \prod_{x} d\phi(x)  d\bar\phi(x) dH(x) \, 
d\chi(x) d\eta(x) \prod_\mu dU_\mu(x)d\psi'_\mu(x) \nonumber \\
& = & \prod_{x} d\phi(x)  d\bar\phi(x) dH(x) \, 
d\chi(x) d\eta(x) \prod_\mu dU_\mu(x)d\psi_\mu(x) \times \det(\cR_\mu)^{-1}, 
\label{dmu_final} 
\eea
where $dU_\mu(x)$ denotes the $SU(N)$ Haar measure, and the measure of the adjoint fields 
	$F(x)=\sum_\alpha F_\alpha(x) T_\alpha$ is defined as 
 $dF(x) \equiv \prod_{\alpha}  dF_\alpha(x)$. 
The expression $d\phi(x) d\bar\phi(x)$ means that the usual measure is used for the real and imaginary parts of the complex field $\phi(x)$. 
The $Q^{imp}$-invariance of (\ref{dmu_final}) follows from the fact that the natural measure (\ref{naive measure}) is invariant 
under (\ref{naive_QU})-(\ref{naive_Qchi})~\cite{Sugino:2006uf}. 

In conclusion, we can use any of them. 
In fact, the difference of the two measures is only the factor $\det(\cR_\mu)^{-1}$
behaving as ${\it const.} \times (1 + {\cal O}(e^{-\ell/a}))$ ($\ell=La$ is the physical size of the system) 
in the continuum limit 
at least for $\cR_\mu^{(\pm)}$ given in  (\ref{Rmu+_r}) and (\ref{Rmu-_r}), 
as we will show in section~\ref{sec:correct_measure}. 
Therefore, even if the natural measure (\ref{naive measure}) breaks the 
$Q^{imp}$ symmetry explicitly, it is negligible in the continuum limit. 
In the same way, the factor $\det(\cR_\mu)^{-1}$ in the $Q^{imp}$-invariant measure
(\ref{dmu_final}) does not affect the continuum limit of the improved theory.

Here we make a comment that there is an interesting exception: 
For $\cR_\mu^{(\e)}$ given in (\ref{Rmue}), 
\eqref{naive measure} is identical with \eqref{dmu_final} 
since $\det (\cR_\mu^{(\e)})=1$ as seen in the last paragraph in appendix~\ref{app:Rmu}. 
Then, 
the natural measure \eqref{naive measure} is also $Q^{imp}$-invariant.

\subsection{Convenient expressions in numerical simulations}
	\label{sec:convenient for simulation}

The standard Monte-Carlo simulation is applicable for the present improved action 
because the boson action is semi-positive definite. 
Of course, it is for the case that the same $\cR_\mu$ is chosen in (\ref{Xi_lat_imp}) and (\ref{QU_final}). 
However the action with (\ref{R_mu_final}) has  the exponentially local interactions and the actual numerical computations 
would demand considerable tasks. 

We can avoid this difficulty by choosing such $\cR_\mu$ that its inverse 
becomes ultra-local like
\begin{equation}
\cR_\mu =\cR^{(-)}_\mu\equiv 
\left(1 - \frac{a}{2}\nabla^*_\mu -r a^2 \nabla_\mu^* \nabla_\mu\right)^{-1} \qquad (r>0), 
\label{R_local_action}
\end{equation}
as well as the ultra-local $\cR_{12}$ given in (\ref{R12_final}).~\footnote{
Here and in (\ref{Xi_lat_imp1}), we can also use more general (\ref{R12-}) with $r>\frac14$ as $\cR_{12}$.
}
Then, in addition to the ultra-local transformations (\ref{QU_final})-(\ref{Qchi_final}), the integrand 
$\Xi_{\rm lat}^{imp}$ becomes also ultra-local:
\begin{equation}
\Xi_{\rm lat}^{imp}
= \tr \biggl\{ \frac{1}{4} \eta [\phi,\bar\phi]
 -i \sum_{\mu=1}^2 (\cR_\mu^{-1}\psi'_\mu) \,(\cR_\mu^{-1} \nabla_\mu \bar\phi)
+\chi \left( H - \frac{i}{a^2} \cR_{12}\Phi_{\rm TL}) \right) \biggr\},
\label{Xi_lat_imp0}
\end{equation}

We may use these field variables $\psi'_\mu$ with (\ref{R_local_action}) and \eqref{dmu_final}
to define the improved theory used in the Monte-Carlo simulations. 
However, we must use $\psi_\mu$ to define observables because the tree-level $\cO(a)$ improvement is achieved for $\psi_\mu$ rather than $\psi'_\mu$. 

Somewhat surprisingly, instead of (\ref{R_local_action}),  if we take 
\begin{equation}
\cR_\mu =\cR^{(\e)}_\mu\equiv \exp \left(\frac{a}{2}\nabla_\mu^{(s)}\right),
\label{R_exp}
\end{equation}
where $\nabla_\mu^{(s)}=\frac{1}{2}(\nabla_\mu+\nabla_\mu^*)$, 
the situation becomes much simpler thanks to the property $\cR_\mu^T = \cR_\mu^{-1}$. 
The integrand $\Xi_{\rm lat}^{imp}$ becomes 
\begin{equation}
\Xi_{\rm lat}^{imp}
= \tr \biggl\{ \frac{1}{4} \eta [\phi,\bar\phi]
 -i \sum_{\mu=1}^2 \psi'_\mu \, \nabla_\mu \bar\phi
+\chi \left( H - \frac{i}{a^2} \cR_{12}\Phi_{\rm TL} \right) \biggr\}
\label{Xi_lat_imp1}
\end{equation}
with (\ref{R12_final}). 
Of course, also in this case, the improved lattice action is given by an ultra-local form.
 The $Q^{imp}$-transformations are the same with those of the unimproved theory 
 and the integrand (\ref{Xi_lat_imp1}) is also mostly the same with
the original one. The $\cO(a)$ improvement is then encoded only in (\ref{psi-prime}) and the definition of the improved lattice field tensor $\Phi^{imp}_{\rm TL}=\cR_{12}\Phi_{\rm TL}$.

In appendix~\ref{app:action}, we present the explicit form of the lattice action 
obtained by acting $Q^{imp}$ to $\Xi^{imp}_{\rm lat}$.  

\subsection{Absence of the fermion doublers}
The doubler modes do not exist in the bosonic sector of the improved theory because the kinetic terms for the scalar and the gauge fields are defined by the forward difference operator with invertible $\cR_\mu$ and $\cR_{12}$.
The exact supersymmetry implies that the fermionic sector should also be free from the doubling problem. 
We will show it by examining the free lattice Dirac operator, explicitly.

The fermion kinetic terms of the improved free theory 
for $\Phi_1$ (\ref{Phi_admissibility}) and $\Phi_2$ (\ref{Phi_tanM})
are given by
\be
S^{imp}_{\rm lat,\, F} = \frac{a^2}{2g^2}\sum_{x\in \Lambda_L} \tr\left[i \sum_{\mu=1}^2 \psi_\mu(x) {\tilde \Delta}_\mu\eta(x) 
+2i {\tilde \chi}(x) 
\left({\tilde \Delta}_1\psi_2(x)- {\tilde \Delta}_2\psi_1(x)\right)\right],
\label{S_fkin}
\ee
where
\be
 {\tilde\Delta}_\mu\equiv \left(\cR^{(0)}_\mu\right)^{-1} \Delta_\mu, \qquad 
 {\tilde \chi} (x) \equiv (\cR^{(0)}_1\cR^{(0)}_2\cR^{(0)}_{12})^T\chi(x),
\label{redefinition}
\ee 
with the forward and backward difference operators,
\begin{eqnarray}
&& \Delta_\mu f(x) \equiv \frac{1}{a}\left(f(x+a\hat{\mu})-f(x)\right), \quad
\label{fwd_op}
\\
&& \Delta^*_\mu f(x) \equiv \frac{1}{a}\left(f(x)-f(x-a\hat{\mu})\right).
\label{bwd_op}
\end{eqnarray}
$\cR_\mu^{(0)}$ and $\cR_{12}^{(0)}$ denote $\cR_\mu$ and $\cR_{12}$ with the gauge field turned off, respectively. 
Namely, for simple examples of $\cR_\mu$ (\ref{Rmu+_r})-(\ref{Rmue}) and $\cR_{12}$ (\ref{R12+})-(\ref{R12e}),  the covariant operators $\nabla_\mu$ and $\nabla_\mu^*$ there are replaced by (\ref{fwd_op}) and (\ref{bwd_op}), 
respectively.~\footnote{
In $\cR^{(0)}_\mu$ and $\cR^{(0)}_{12}$, the transpose operation 
denoted by the superscript $T$ maps $\Delta_\mu$ and $\Delta_\mu^*$ 
to  $-\Delta^*_\mu$ and $-\Delta_\mu$, respectively.
}
Note that $\cR^{(0)}_\mu$ and $\cR^{(0)}_{12}$ commute with each other
since they do not depend on the gauge field.

The form of (\ref{S_fkin}) is the same as that of the original unimproved lattice model~\cite{Sugino:2003yb,Sugino:2004qd} 
which has no doubler modes, except $\cR^{(0)}_\mu$ and $\cR^{(0)}_{12}$ are included in the redefinition (\ref{redefinition}). 
It is clear that the improved theory also has no doublers
since the improvement factors 
in (\ref{redefinition}) are invertible and do not affect 
the conclusion in the original model.

Explicitly, in terms of the four-component spinor $\tilde{\Psi}\equiv \left(\psi_1,\psi_2,\tilde{\chi},\frac12\eta\right)^T$, (\ref{S_fkin}) is expressed as~\footnote{The transpose operation of $\Psi(x)$ acts only to the spinor indices.}  
\be
 S^{imp}_{\rm lat,\, F} = \frac{a^2}{2g^2}\sum_{x\in \Lambda_L} \tr\left[\tilde{\Psi}(x)^T\tilde{D}\tilde{\Psi}(x)\right],
\ee
where
\be
 \tilde{D} \equiv \sum_{\mu=1}^2\left[-\frac12\gamma_\mu\left(\tilde{\Delta}_\mu-\tilde{\Delta}_\mu^T\right)
-\frac12P_\mu \left(\tilde{\Delta}_\mu+\tilde{\Delta}_\mu^T\right)\right],
\label{Dirac_op}
\ee
with 
\bea
& & \gamma_1=-i\sigma_1\otimes\sigma_1, \qquad \gamma_2=i\sigma_1\otimes\sigma_3, \nn \\
& & P_1=\sigma_1\otimes\sigma_2,\qquad P_2=\sigma_2\otimes \id_2
\eea
satisfying 
\be
\{\gamma_\mu,\,\gamma_\nu\}=-2\delta_{\mu\nu}, \qquad \{P_\mu,\,P_\nu\}=2\delta_{\mu\nu},\qquad 
\{\gamma_\mu,\,P_\nu\}=0.
\label{gamma_P}
\ee
Since the Dirac operator (\ref{Dirac_op}) is Hermitian, we may consider the zero of $\tilde{D}^2$ in order to see that of $\tilde{D}$. From~(\ref{gamma_P}),  
$
\tilde{D}^2=\sum_{\mu=1}^2\tilde{\Delta}_\mu\tilde{\Delta}_\mu^T,
$
which means that $\tilde{D}$ vanishes only at the zero of $\tilde{\Delta}_\mu$ that is nothing but the zero of $\Delta_\mu$ (the origin in the momentum space) since $\cR_\mu$ are invertible. 

In the ordinary Wilson-Dirac operator, fermion doublers are lift by the Wilson term of the order of $\cO(a)$. 
Since our lattice action is improved and has no $\cO(a)$ term, 
the Wilson term appearing in (\ref{Dirac_op}) should be higher order in $a$.~\footnote{
The factor appearing in the redefinition of $\chi$~(\ref{redefinition}) does not affect the $\cO(a)$ contribution since it behaves as 
$1+\cO(a^2)$ for any $\cR_\mu$ and $\cR_{12}$.
Also, it does not lift the doublers in the lattice Dirac operator $D$ which is defined by
 $S^{imp}_{\rm lat, F}=\frac{a^2}{2g^2} \sum_{x\in \Gamma_L} \tr (\Psi^T D\Psi)$
where $\Psi=\left(\psi_1,\psi_2,\chi, \frac{1}{2}\eta \right)^T$. 
} 
Let us see this in the case 
of $\left.\cR_\mu^{(-)}\right|_{r=\frac12}$ in (\ref{Rmu-_r}). 
The Dirac operator can be written as 
\begin{align}
\tilde{D} = & -\frac12\sum_\mu\gamma_\mu(\Delta_\mu+\Delta_\mu^*)  -\frac{a}{2}\sum_\mu P_\mu(\Delta_\mu\Delta_\mu^*)  \nn \\
&+\frac{a^2}{4}\sum_\mu\gamma_\mu\left(\Delta_\mu + \Delta_\mu^*\right)(\Delta_\mu\Delta_\mu^*) +\frac{a}{4}\sum_\mu P_\mu\left((\Delta_\mu)^2+(\Delta_\mu^*)^2\right). 
\label{Dirac_op_a2}
\end{align}
The first and the second terms are the standard kinetic term and the Wilson term, respectively, in the original unimproved model,  
while the third and fourth terms are generated by the $\cO(a)$ improvement.
The first term reproduces the naive kinetic term in the continuum limit up to $\cO(a)$, and the third term has the same zeros with those of the first term.
Interestingly, 
the second and fourth terms combine to yield the $\cO(a^3)$ Wilson term $\frac{a^3}{4}\sum_{\mu}P_\mu(\Delta_\mu\Delta_\mu^*)^2$. 
Thus, we have 
\begin{align}
\tilde{D} = & -\frac12\sum_\mu\gamma_\mu(\Delta_\mu+\Delta_\mu^*)\left(1-\frac{a^2}{2}\Delta_\mu\Delta_\mu^*\right)  
+\frac{a^3}{4}\sum_\mu P_\mu(\Delta_\mu\Delta_\mu^*)^2,
\label{Dirac_op_a}
\end{align}
with the second term being the Wilson term in the improved action actually of the order of $a^3$. 

\subsection{Path-integral measure}
\label{sec:correct_measure}

In section \ref{subsec:result}, we have defined the path-integral measure (\ref{dmu_final}) 
which is invariant under the improved transformations (\ref{QU_final})-(\ref{Qchi_final}). 
In what follows, we show that the factor $\det \cR_\mu$ is irrelevant 
in the continuum limit $a\to 0$ with keeping the physical length $\ell\equiv La$ fixed, and 
the natural measure is therefore reproduced 
without breaking the tree-level $\cO(a)$ improvement,
at least, 
for  $\cR_\mu^{(+)}$ 
and  $\cR_\mu^{(-)}$ given in (\ref{Rmu+_r})  and (\ref{Rmu-_r})
as announced.

We only have to evaluate the determinant of $R_\mu^{(+)}$ because one can show that
\be
\det\,\cR_\mu^{(-)}=\left(\det\,\cR_\mu^{(+)}\right)^{-1},
\ee
as mentioned in appendix \ref{app:Rmu}. 
For simplicity of the explanation, we focus on the case of $\det \cR_1^{(+)}$. 
Of course, the same result is obtained for $\cR_2^{(+)}$ by interchanging the role of the directions $1$ and $2$.

The matrix representation of $\cR^{(+)}_\mu$ can be extracted as (\ref{eq:kernel_R}) for $x,y\in \Lambda_L$ and $\alpha,\beta=1,2,\cdots,N^2-1$:
\bea
(\cR_\mu^{(+)})_{\alpha\beta}(x,\,y)  
& =  &\delta_{x_\rho,\,y_\rho}\Biggl[\left(\frac12+2r\right)\delta_{x_\mu,\,y_\mu}\delta_{\alpha\beta}
+\left(\frac12-r\right)
\delta_{x_\mu+a,y_\mu}\,
\hat{U}_{\mu,\alpha\beta}(x)   \nn \\
& & \hspace{12mm}
-r\delta_{x_\mu-a,y_\mu}\,\hat{U}_{\mu,\beta\alpha}(y)\Biggr]. 
\label{Rmu+measure}
\eea
Here $\rho$ denotes the direction other than $\mu$ ($\rho=1$ and 2, when $\mu=2$ and 1, respectively), 
and $\hat{U}_{\mu,\alpha\beta}(x)$ is the adjoint representation 
of $U_\mu(x)$:
\be
\hat{U}_{\mu,\alpha\beta}(x) \equiv \tr\left[T_\alpha U_\mu(x)T_\beta U_\mu(x)^{-1}\right].
\label{Fmu}
\ee
Likewise, we express the adjoint representation of an $SU(N)$ 
matrix $A$
by putting a hat as 
\begin{equation}
	\hat{A}_{\alpha\beta} \equiv \tr\left[ 
	T_\alpha\,A\,T_\beta\,A^{-1} \right].
	\label{fund-adj}
\end{equation}
Note that $(\hat{A}^{-1})_{\alpha\beta}=\hat{A}_{\beta\alpha}$ is 
always satisfied.

We should note that (\ref{Rmu+measure}) is an ${\cal M}\times {\cal M}$ matrix (${\cal M}\equiv L^2(N^2-1)$) which is diagonal with respect to the $\rho$ direction. 
In order to evaluate $\det \,\cR_1^{(+)}$ explicitly, 
we first diagonalize (\ref{Rmu+measure}) in the color space. 
To this end, let us consider a gauge function, 
\begin{equation}
	g_1(x_1,x_2) \equiv 
	\begin{cases}
		1, & (x_1=La) \\
		\left( P_1(x_2) \right)^{-\frac{x_1}{La}} U_1(0,x_2) \cdots U_1(x_1-a,x_2), & (x_1=a,2a,\cdots,(L-1)a)
	\end{cases}
\end{equation}
where $P_1(x_2)$ is the Polyakov line along the $x_1$-direction:
\begin{equation}
	P_1(x_2) \equiv U_1(0,x_2)U_1(a,x_2)\cdots U_1((L-1)a,x_2). 
\end{equation}
Then we can eliminate the $x_1$ dependence of $U_1(x_1,x_2)$ by 
the gauge transformation with $g_1(x_1,x_2)$ as  
\begin{equation}
    g_1(x_1,x_2) U_1(x_1,x_2) g_1(x_1+a,x_2)^{-1} =P_1(x_2)^{1/L}. 
\end{equation}
This means that $\hat{U}_1(x_1,x_2)$ 
in (\ref{Rmu+measure}) is given by 
a gauge transformation of (the $L$-th root of) the Polyakov line 
in the adjoint representation: 
\begin{equation}
	\hat{U}_1(x_1,x_2) =  
	\hat{g}_1(x_1,x_2)^{-1} 
	\hat{P}_1(x_2)^{1/L} 
	\hat{g}_1(x_1+a,x_2).
	\label{PtoU}
\end{equation}
Recall that the hatted variables mean the adjoint representation of 
the corresponding unitary matrices as mentioned around (\ref{fund-adj}).

The eigenvalues of the adjoint Polyakov line $\hat P_1(x_2)$ are given by
\begin{equation}
p_{1}(x_2,\alpha) \in S^1, \qquad {\rm for} \ \  \alpha=1,\cdots,N^2-1.
\label{ev_of_P}
\end{equation}
Then, $\hat{P}_1(x_2)$ can be diagonalized as 
\begin{equation}
	\hat{P}_1(x_2) =W_1(x_2)^{-1} \textrm{diag}\left\{ p_{1}(x_2,1),p_{1}(x_2,2)\cdots,
	p_{1}(x_2,N^2-1)
	\right\}  W_1(x_2),
	\label{diagonalize}
\end{equation}
where $W_1(x_2)$ is a unitary matrix with the size of $N^2-1$.

Combining (\ref{PtoU}) and (\ref{diagonalize}) into $V_{\alpha\beta}(x,y)=\delta_{x_1,y_1}\delta_{x_2,y_2} (W_1(x_2) \hat g_1(x_1,x_2) )_{\alpha\beta}$,
we can diagonalize 
$\cR_1^{(+)}$ in (\ref{Rmu+measure}) with respect to the color index $\alpha$:
\begin{equation}
	(V\cR_1^{(+)}V^{-1})_{\alpha\beta}(x,y) = \delta_{x_2,y_2}\delta_{\alpha\beta}
	\cD_{x_2,\alpha}({x_1,y_1}), 
\end{equation}
where 
\begin{align}
	\cD_{x_2,\alpha}({x_1,y_1})
	= 
		A\delta_{x_1,y_1} 
		+B \delta_{x_1+a,y_1}
		+C \delta_{x_1-a,y_1},
		\label{R1+_diagonal}
\end{align}
with
\begin{equation}
A=\frac12+2r, \quad 
B=\left(\frac12-r\right) ({p}_1(x_2,\alpha))^{\frac{1}{L}}, \quad 
C=-r ({p}_1(x_2,\alpha))^{-\frac{1}{L}}.
\label{ABC_R1+tilde}
\end{equation}

 Now, the computation of $\det \cR_1^{(+)}$ reduces to evaluating the determinant of  
$L\times L$ matrix $\cD_{x_2,\alpha}$: 
\be
\det \,\cR_1^{(+)} = \prod_{x_2=a}^{La}\prod_{\alpha=1}^{N^2-1} 
\det\left(\cD_{x_2,\,\alpha}\right),
\label{Det_R1+}
\ee
where
$\cD_{x_2,\alpha}$ in  (\ref{R1+_diagonal}) is expressed in the form of 
\begin{align}
	\cD_{x_2,\alpha}
=\begin{pmatrix} A & B&   &  C \\
C & \ddots &  \ddots&     \\
& \ddots & \ddots  &  B  \\
B  &  &   C  & A    
\end{pmatrix}
.
\label{R1+_matrix}
\end{align}
The determinant of the circulant matrix (\ref{R1+_matrix}) can be evaluated 
straightforwardly (see appendix~\ref{app:determinant}) and the result is 
\be
\det(\cD_{x_2,\alpha}) =  \xi_+^L + \xi_-^L -(-B)^L-(-C)^L, 
\label{det_D_solution}
\ee
where $\xi_\pm=\frac12\left(A\pm \sqrt{A^2-4BC}\right)$.

We thus find that
\begin{align}
\det \,\cR_1^{(+)} = \prod_{x_2=a}^{La}\prod_{\alpha=1}^{N^2-1} 
 \left( F(r,L) - G(r,L,p_1(x_2,\alpha) \right), 
\label{detRL_sol}
\end{align}
where 
\begin{align}
	F(r,L) &\equiv \left(\frac{1+4r+\sqrt{1+16r}}{4}\right)^L
	+\left(\frac{1+4r-\sqrt{1+16r}}{4} \right)^L, \\ 
	G(r,L,p) &\equiv 
	 \left(r-\frac{1}{2} \right)^L p
	+ r ^L p^{-1}. 
\end{align}
$G(r,L,p_1(x_2,\alpha)) $ 
carry information of the gauge fields via the phase factor $p_1(x_2, \alpha)$ (\ref{ev_of_P})
whereas $F(r,L)$ does not.
We can say that $\det\,\cR_1^{(+)}$ is irrelevant in the continuum limit
if $G$ 
becomes negligible compared with $F$ 
as $a\rightarrow 0$ 
(keeping $\ell=La$ fixed). 
It is straightforward to see that 
$G/F=\cO(e^{-l/a})$ as $a\rightarrow 0$
 for $r>0$. 
We can repeat the same argument for $\det\,\cR_2^{(+)}$ and obtain the same conclusion. 

Thus, we reach the final point: {\em For $r>0$, the Jacobian factor $\prod_\mu\det\,\cR_\mu^{(\pm)}$ is irrelevant in the continuum limit, 
and both of the measures (\ref{naive measure}) and (\ref{dmu_final}) can be used to define  the O(a)-improved theory.
}\footnote{
If radiative corrections are taken into account, the speed of the convergence to the continuum limit could be different. 
}

\section{Conclusion and discussions}
\label{sec:conclusion}
In this paper we have discussed tree-level ${\cal O}(a)$ improvement of a lattice formulation of 
2d ${\cal N}=(2,2)$ super Yang-Mills theory 
introduced by one of the authors (F.~S.)~\cite{Sugino:2003yb}.  
Technically important ingredient is the improvement of the $Q$-transformation, 
the term $\Xi$ in \eqref{S_lat} and the path-integral measure. 
The problem of zero-modes arising in improving the $Q$-transformation is resolved 
by the use of a kind of the Wilson terms, which leads to the action containing exponentially local terms in general. 

In the framework of the Symanzik $\cO(a)$ improvement~\cite{Symanzik:1979ph,Weisz:2010nr}, 
we should also consider the improvement of the effective action at the loop level. 
Namely, we have to determine $\cO(a)$ counterterms to be added to the lattice action so that 
they cancel all of $\cO(a)$ radiative corrections. 
Thanks to the superrenormalizable property of the theory, investigation at the one-loop level is sufficient.   
This will be reported in the forthcoming publication. 

Needless to say, the most important application of this method is the actual numerical simulations.
It is interesting to see how the tree-level improvement discussed here 
accelerates the simulation and how the addition of the one-loop improvement changes the situation.

The same idea can be applied to other theories, with different amount of supersymmetries and/or 
with various matter fields. 
In particular, the application to a similar lattice formulation of 2d $\cN=(4,4), (8,8)$ super Yang-Mills theories 
that preserves two supercharges~\cite{Sugino:2003yb,Sugino:2004qd} should be straightforward. 
It would be interesting to consider if the improvement can be generalized to other types of supersymmetric 
lattice formulations~\cite{Kaplan:2002wv,Catterall:2003wd,Cohen:2003xe,DAdda:2004jb}. 
To a plane wave deformation of 2d ${\cal N}=(8,8)$ super Yang-Mills theory on lattice~\cite{Hanada:2010kt,Hanada:2010gs}, 
from which 4d ${\cal N}=4$ SYM can be obtained without parameter fine tunings, the application of the improvement 
is worth being investigated.

\section*{Acknowledgement}
This work is supported by JSPS KAKENHI Grant Number JP16K05328 and the MEXT-Supported Program for the Strategic
Research Foundation at Private Universities Topological Science (Grant No. S1511006). 
The work of M.~H. is partly supported by Grant-in-Aid for Scientific Research JP25287046 and JP17K14285.
The work of S.~M. is partly supported by Grant-in-Aid for Scientific Research (C) 15K05060.
The work of F.~S. is partly supported by Grant-in-Aid for Scientific Research (C), 25400289, and by IBS-R018-D2.

\appendix
\section{Proof of Lemma}\label{app:lemma}
In this appendix, we prove the lemma in section~\ref{subsec:SUSY improve}. 
Since the essential points are not lost by suppressing Lorentz and sites indices, we show it without them:
\begin{lem}
Let $f$ be a function that depends on $(U,\psi,\phi)$ with the same gauge transformation property as $U$,  and suppose 
that
\be
QU= f(U, \psi, \phi). 
\label{app:lemma_QU}
\ee
Let $g$  be a function that depends on $(U,QU,\phi)$  with the same gauge transformation property as $\psi$. 
If (\ref{app:lemma_QU}) can be solved in term of $\psi$ as follows: 
\be
\psi = g(U, QU,\phi),
\label{app:lemma_psi}
\ee 
then one can consistently define the $Q$-transformation by (\ref{app:lemma_QU}) and 
\bea
Q\psi & = & \left[QU\cdot \frac{\delta}{\delta U} -i\delta_\phi U\cdot \frac{\delta}{\delta(QU)}\right]
g (U, QU,\phi),
\label{app:lemma_Qpsi} \\
Q\phi & = & 0,
\label{app:lemma_Qphi} 
\eea 
so that $Q^2=-i\delta_\phi$ for all fields.
\end{lem}

\noindent {\it Proof}:
We immediately confirm that $Q^2=-i\delta_\phi$ for $\phi$ since the both sides vanish from (\ref{app:lemma_Qphi}) and the gauge transformation $\delta_\phi\phi=0$. The main task is to show the lemma for $U$ and $\psi$. 

Acting $Q$ to (\ref{app:lemma_QU}) leads to
\be
Q^2U= \left[QU\cdot \frac{\delta}{\delta U}+Q\psi\cdot \frac{\delta}{\delta\psi}\right]f(U, \,\psi, \,\phi).
\label{app:lemma2}
\ee
Plugging (\ref{app:lemma_Qpsi}) into (\ref{app:lemma2}), 
we have 
\be
Q^2U=-i \delta_\phi U \cdot \left\{ \frac{\delta}{\delta(QU)}g \cdot \frac{\delta}{\delta\psi} f \right\}
 +QU\cdot \left\{\frac{\delta}{\delta U}f
+ \frac{\delta}{\delta U}g \cdot \frac{\delta}{\delta\psi} f\right\}.
\label{app:lemma3}
\ee
Once the variable $\psi$ in $f$ is eliminated by using (\ref{app:lemma_psi}),
$f$ is a function of $U,QU,\phi$ as $f(U,g(U,QU,\phi),\phi)$
which does not actually depend on $U$ and $\phi$ because of (\ref{app:lemma_QU}).
It means that, in the r.h.s. of (\ref{app:lemma3}),  
the expression inside of  
the first curly bracket is one
and the second one vanishes. Thus, we obtain 
\be
Q^2U=-i\delta_\phi U.
\label{app:lemma4}
\ee
By operating a bosonic transformation $Q^2$ to (\ref{app:lemma_psi}), we find 
\be
Q^2\psi=\left[Q^2U\cdot \frac{\delta}{\delta U} +Q^3 U\cdot \frac{\delta}{\delta(QU)}\right]
g(U, \,QU, \,\phi). 
\ee
Note that $Q^3U=-iQ\delta_\phi U=-i\delta_\phi (QU)$ from (\ref{app:lemma4}).   It is found that 
\be
Q^2\psi=-i\delta_\phi \psi.
\label{app:lemma5}
\ee
(\ref{app:lemma4}) and (\ref{app:lemma5}) establish the statement. \qed

The remarkable point of the proof above is that it relies only on the algebraic structure of $Q$-transformation. 
This kind of argument is, therefore, applicable beyond the framework of the lattice gauge theory.
In particular, $U$ is not needed to be a unitary variable, and $x$-space is not limited to the lattice.
If the gauge symmetry is realized in a framework, 
this lemma allows us to construct the $Q$-transformation satisfying $Q^2 = -i \delta_\phi $  in it.

\section{$\cR_\mu$ and $\cR_{12}$}\label{app:R}
We present several $\cR_\mu$ and  $\cR_{12}$ which can be used for improving the lattice $Q$-transformation and the lattice field tensor, respectively.

\subsection{$\cR_\mu$}\label{app:Rmu}
In sections \ref{subsec:SUSY improve} and \ref{sec:correct_measure}, we have explained that $\cR_\mu$ should obey several conditions: 
it behaves as $1+\frac{a}{2}D_\mu$ near the continuum limit, and both of $\cR_\mu$ and its inverse $\cR_\mu^{-1}$ are local. 
In addition, the gauge-field dependence of ${\det (\cR_\mu) }$ rapidly decays when taking the continuum limit. 

There are actually infinite solutions for $\cR_\mu$. 
We list a few types of $\cR_\mu$ that satisfy the conditions above:
\begin{eqnarray}
&&\cR_\mu^{(+)} \equiv 1 + \frac{a}{2}\nabla_\mu -r a^2 \nabla_\mu^* \nabla_\mu \qquad (r>0) ,
\label{Rmu+_r}
\\
&&\cR^{(-)}_\mu \equiv 
\left(1 - \frac{a}{2}\nabla^*_\mu -r a^2 \nabla_\mu^* \nabla_\mu\right)^{-1} \qquad (r>0), 
\label{Rmu-_r}
\\
&&\cR^{(\e)}_\mu\equiv \exp\left(\frac{a}{2}\nabla_\mu^{(s)}\right),
\label{Rmue}
\end{eqnarray}
where $\nabla_\mu^{(s)}$ is the covariant symmetric difference operator,
\be
\nabla_\mu^{(s)}\equiv \frac12\left(\nabla_\mu + \nabla_\mu^*\right).
\label{symmetric_diff}
\ee
These operators have different properties as explained below.  

The first one (\ref{Rmu+_r}) is 
the simplest ultra-local solution which is given in section  \ref{subsec:SUSY improve}. 
For 
$r=\frac14$ and $r=\frac12$, (\ref{Rmu+_r}) becomes further simple depending only on the symmetric and the backward 
difference operators linearly, respectively. 
From the fact that the eigenvalue of $a\nabla_\mu$ is expressed in the form (\ref{nabla_EV}), 
$\cR_\mu^{(+)}$ is bounded in the sense that the absolute value of the singular values has the upper and non-zero lower bounds. Namely, 
\be
u \le \left(\cR_\mu^{(+)}\right)^\dag \cR_\mu^{(+)} \le v, \qquad  {\rm for \ } u,v>0,
\label{bounded}
\ee
where the inequality is understood as for the eigenvalues of $\cR^\dag \cR$.
Lemma 2 given in section \ref{app:locality} tells us that the ultra-local and bounded operator has a local inversion with an exponentially decaying tail:
\be
\left| (\cR^{(+)}_\mu)^{-1} (x,y)\right| 
\leq C\,e^{-\rho \,|x_\mu-y_\mu|/a},
\label{locality_R+}
\ee
where $C$ and $\rho$ are positive constants. So, we can make use of this operator with keeping the principle of locality. 
As shown in section \ref{sec:correct_measure},  ${\det (\cR_\mu^{(+)})}$ is $Q$-invariant up to a strongly suppressed 
term in the continuum limit ${\rm exp}(-\l/a)$ (with the physical size $\l$ is fixed in the limit).

In this way, the operator (\ref{Rmu+_r}) correctly improves the $Q$-transformation and is useful to explain the method of the tree-level ${\cal O}(a)$ improvement. 
However, the numerical application will not be easy since the lattice action has an exponentially local interactions
because $Q\psi_\mu$ yields the factor $(\cR_\mu^{(+)})^{-1}$ that spreads all over the lattice.

Instead,  the second one (\ref{Rmu-_r})   provides an ultra-local action which is suitable for numerical applications. 
As discussed in section~\ref{sec:convenient for simulation}, with the change of variables in the fermion fields $\psi_\mu$, the lattice action is actually expressed as  an ultra-local expression.
This is because $(\cR^{(-)}_\mu)^{-1}$ is taken to be ultra-local. 
Since  $(\cR_\mu^{(-)})^{-1}=(\cR_\mu^{(+)})^{T}$ is derived from  $\nabla_\mu^T = -\nabla_\mu^*$, 
  $\cR^{(-)}_\mu$ satisfies the exponentially locality condition such as (\ref{locality_R+}).   
As seen in section~\ref{sec:correct_measure}, the fermion measure is also consistently defined as long as $r>0$.   
In this case, 
we also obtain simple expressions for $r=\frac14$ and $r=\frac12$.  
The latter is convenient in writing the improved action explicitly. 

The third choice $\cR^{(\e)}_\mu$ looks complicated for numerical applications but is worth noting since it has a trivial determinant:
\be
\det \,\cR_\mu^{(\e)}=1, 
\label{detR}
\ee
that is, the measures (\ref{naive measure}) and (\ref{dmu_final}) are identical and $Q^{imp}$-invariant. 
(\ref{detR}) is easily shown from $(\nabla_\mu^{(s)})^T=-\nabla_\mu^{(s)}$, or equivalently, 
from the fact that  $\cR_\mu^{(\e)}$ is an orthogonal matrix  whose determinant is $+1$ or $-1$. The sign is fixed because (\ref{detR}) holds in the free limit and 
the sign does not change under the continuous deformation of the link fields. 
Lemma 3  in appendix~\ref{app:locality} allows us to conclude that 
\be
\left| \cR^{(\e)}_\mu (x,y)\right| 
\leq \frac{C\,e^{-\rho \,|x_\mu-y_\mu|/a}}{\left(|x_\mu-y_\mu|/a\right)!},
\ee
with positive constants $C$ and $\rho=\ln 2$. 
Due to the factorial growth of the denominator, $\cR_\mu^{(\e)}$ decays much faster than general exponentially local 
operators as $|x_\mu-y_\mu|/a\to \infty$.

\subsection{$\cR_{12}$}\label{app:R12}  
  
The lattice field tensor is improved by multiplying $\cR_{12}$ which obeys the several conditions as with $\cR_\mu$. 
As seen in section \ref{subsec:differece improve},  $\cR_{12}$ should behaves as $1-\frac{a}{2}(D_1+D_2)$ near the continuum limit, 
and should be local and invertible to lead to the correct continuum theory around the unique vacuum.
There are also infinite possibilities satisfying these conditions.

We give a few $\cR_{12}$ which are similar to $\cR_\mu$ (\ref{Rmu+_r})-(\ref{Rmue}):
\begin{eqnarray}
&& \cR_{12}^{(+)}\equiv \left[1+\frac{a}{2}\sum_{\mu=1}^2\nabla_\mu -ra^2\sum_{\mu=1}^2\nabla_\mu^*\nabla_\mu\right]^{-1},
\label{R12+}
\\
&& \cR_{12}^{(-)}\equiv 1-\frac{a}{2}\sum_{\mu=1}^2\nabla_\mu^* -ra^2\sum_{\mu=1}^2\nabla_\mu^*\nabla_\mu,
\label{R12-}
\\
&&\cR_{12}^{(\e)}\equiv \exp\left(-\frac{a}{2}\sum_{\mu=1}^2\nabla_\mu^{(s)}\right).
\label{R12e}
\end{eqnarray}
These operators coincide with each other up to ${\cO}(a)$, that is, they improve $\Phi_{\rm TL}$ to the order as shown in section \ref{subsec:differece improve}.~\footnote{
These operators maintain the reflection symmetry of the original lattice action (\ref{S_lat}): 
$x=(x_1,\,x_2)\to \tilde{x}\equiv(x_2,\,x_1)$ with 
\bea
(U_1(x),\,U_2(x)) & \to & (U_2(\tilde{x}), U_1(\tilde{x})) \nn \\
(\psi_1(x),\,\psi_2(x)) & \to & (\psi_2(\tilde{x}), \psi_1(\tilde{x})) \nn \\
(H(x), \chi(x)) & \to & (-H(\tilde{x}), -\chi(\tilde{x})) \nn \\
(\phi(x), \bar{\phi}(x), \eta(x)) & \to & (\phi(\tilde{x}), \bar{\phi}(\tilde{x}), \eta(\tilde{x})) .
\eea
}
We can show that (\ref{R12+}) and the inversion of (\ref{R12-}) exist for $r>1/4$ as follows. 
For an operator $Z_\mu$ defined as
\be
Z_\mu \varphi(x) = r U_\mu(x) \varphi(x+a\hat{\mu}) U_\mu(x)^{-1} 
- \left(\frac12 -r\right) U_\mu(x-a\hat{\mu})^{-1} \varphi(x-a\hat{\mu})  U_\mu(x-a\hat{\mu}) 
\ee 
for any adjoint field $\varphi(x)$, 
whose induced norm~\footnote{
$Z_\mu$ and $\varphi$ can be regarded as a matrix of the size $\cM=(N^2-1)L^2$ 
and an $\cM$-dimensional vector, respectively. 
For any non-zero $\cM$-dimensional vector $\vec{u}$ with a norm $||\vec{u}||=\sqrt{\vec{u}^\dagger\vec{u}}$, 
the induced norm of an $\cM\times \cM$ matrix $A$ is defined by 
$
||A||_{\rm ind}\equiv \max_{\vec{u}}\frac{||A\vec{u}||}{||\vec{u}||}$.
}
is less than or equal to $|r| + |\frac{1}{2}-r|$. Then, 
\be
 \cR_{12}^{(-)} = 4r - Z_1 -Z_2
\ee
are shown to have no zeros for $r>1/4$. 
The existence of (\ref{R12+}) obeys $(\cR_{12}^{(+)})^{-1}=(\cR_{12}^{(-)})^T$.
Note that (\ref{R12+}) and (\ref{R12e}) are local operators from Lemmas 2 and 3 given in the next subsection.

For $r=\frac12$, $\cR_{12}^{(\pm)}$ reduce to simpler expressions given with the backward (or the forward) operators,  respectively. In particular, $\cR_{12}^{(-)}$ is convenient to write the ultra-local action with  $\cR_{\mu}^{(-)}$ (\ref{Rmu-_r}) 
as given in appendix \ref{app:action}.

\subsection{Locality of $\cR$}\label{app:locality}
The locality of operators is discussed in detail.  We present the locality conditions for the inverse (the exponential) of an ultra-local operator in Lemma 2 (Lemma 3). 
These give the solid theoretical grounds of the tree-level ${\cal O}(a)$ improvement with $\cR_\mu$ and $\cR_{12}$ in this paper.

For an ultra-local operator $R$, it is easy to show that $R^\dagger$ and $R^\dagger R$ are also ultra-local. 
In Lemma 2, we show that if $R^\dagger R$ has the upper and non-zero lower bounds, $R^{-1}$ satisfies the exponential locality condition.
Although this can be shown by applying an argument in~\cite{Hernandez:1998et}, 
we present a proof to make this paper self-contained as much as possible.

\begin{lem}
If $R$ is an ultra-local operator that satisfies, for  $u,v>0$,
\be
u\leq R^\dagger R \leq v,
\ee
where the inequality stands for the eigenvalues, 
   $R^{-1}$ is exponentially local.
\end{lem}

\noindent {\it Proof:}
 The locality of $R^{-1}$ follows from that of $(R^\dag R)^{-1}$ because
$R^{-1}=(R^\dagger R)^{-1} R^\dagger$.
To study the locality of $(R^\dag R)^{-1}$,  let us set
\be
Z\equiv \frac{2}{u-v}\left(R^\dagger R-\frac{u+v}{2}\right),
\ee 
whose eigenvalues have the absolute value not exceeding one.
Then, with $e^{-\theta}\equiv\frac{v-u}{u+v}<1$, we can show  
\be
(R^\dagger R)^{-1}= \frac{2}{u+v}\sum_{n=0}^\infty \ e^{-n\theta} Z^n, 
\label{RdagR_expansion}
\ee
by expanding $(R^\dagger R)^{-1}=\frac{2}{u+v}\left(1-e^{-\theta}Z\right)^{-1}$ with respect to $Z$.

Since $Z$ is ultra-local from the assumption, there is a positive constant $M$ such that
$(Z^n)(x,y)=0$ for $n< M||x-y||_1/a$.~\footnote{
$M\sim 1/r$ where $r$ is the range of the ultra-locality of $Z$.} 
Thus, 
\be
(R^\dagger R)^{-1}(x,y) = \frac{2}{u+v}\sum_{n=M||x-y||_1/a}^\infty \ e^{-n\theta}  Z^n(x,y). 
\label{app:claim_kernel}
\ee
We can easily show that
\footnote{Let $W$ be an ${\cal M}\times {\cal M}$ matrix whose singular values have the upper bound $w$.
For a norm of complex vectors $||\vec{u}||\equiv\sqrt{\vec{u}^\dagger\vec{u}}$, 
it is found that
$||W \vec{u}||\leq w||\vec{u}||$ holds for any vector $\vec{u}\in \C^{\cal M}$. 
Then, for the matrix elements of $W$, we can easily see
\be
\sqrt{\sum_{I=1}^{\cal M}|W_{IJ}|^2} \leq w\qquad {\rm for \ all \ } J, 
\label{app:claim_W}
\ee
by taking the unit vectors $(\vec{\e}_J)_I\equiv\delta_{IJ}$ as $\vec{u}$, and find that $|W_{IJ}|\leq w$.
}
\be
\left|Z_{\alpha \beta}(x,y)\right| \leq 1, 
\label{app:claim_Zn}
\ee  
and finally obtain
\be
\left|(R^\dagger R)^{-1}_{\alpha\beta}(x,y)\right|
\leq C \ e^{-\rho ||x-y||_1/a},
\ee
where $C$ and $\rho=M\theta$ are constants independent of the lattice spacing.
Namely, $(R^\dagger R)^{-1}$ is exponentially local.  So, we can conclude that $R^{-1}$ satisfies the locality in the same sense. \qed
\vspace{5mm}

The operator $e^R$ is also useful to understand our formulation. The following lemma tells us the condition on $R$ under which $e^R$ is local.

\begin{lem}
If $R$ is an ultra-local operator whose singular values have the upper bound $w>0$, $e^R$ is local in the sense that at long distance it decays faster than the exponential.
\end{lem}

\noindent {\it Proof:} Instead of (\ref{RdagR_expansion}), we have 
\begin{eqnarray}
e^R=\sum_{n=0}^\infty \frac{1}{n!} R^n.
\label{expantion_of_eR}
\end{eqnarray}
Since $(R^n)(x,y)=0$ for $n< M||x-y||_1/a$ for  a positive constant $M$,
the kernel representation of (\ref{expantion_of_eR}) is 
\begin{eqnarray}
e^R(x,y)=\sum_{n=M||x-y||_1/a}^\infty \frac{1}{n!} R^n(x,y).
\end{eqnarray}
Noting that $\left|(R)_{\alpha \beta}(x,y)\right| \leq w$ from the assumption, we obtain 
\begin{eqnarray}
|(e^R)_{\alpha\beta}(x,y)| \le C  \frac{w^{M||x-y||_1/a}}{(M||x-y||_1/a)!},
\end{eqnarray}
where $C$ is a positive constant.
Using an identity ${\rm log}(n!) \ge \kappa n \, {\rm log}\, n $ for $n \gg 1$ with $0 < {}^\exists \kappa<1$, we find that 
\begin{eqnarray}
|(e^R)_{\alpha\beta}(x,y)| \lesssim C' e^{-\kappa \frac{||x-y||_1}{a} {\rm  log}(\frac{||x-y||_1}{a})},
\end{eqnarray}
at large distance $||x-y||_1/a \gg 1$, where $C'$ and $\kappa$ are positive constants.
Since $e^{-n\, {\rm log}\, n}$ is larger than $e^{-n}$ for $n \gg 1$, 
$e^R$ decays faster than the exponential at large distance. \qed

\section{Explicit form of the lattice action}\label{app:action}
In this appendix, we present the explicit form of the lattice action \eqref{improved action} 
after performing $Q^{imp}$-transformations. 
Before seeing it, we also present the case of the unimproved lattice action to clarify the differences 
arising in the improvement.  

The action is divided as 
\be
S  =S_{\rm B} +S_{\rm F}  +S_{\rm Y},
\ee
where $S_{\rm B}$ is the boson action,
and $S_{\rm F}$ and $S_{\rm Y}$ are the fermion actions 
which include kinetic terms and the Yukawa interaction terms, respectively.
In the continuum theory with the twisted variables,  
\bea
&&S_{\rm B}=  \frac{1}{2g^2}\int d^2x\,\tr\biggl\{ H^2 -2i HF_{12} + \sum_{\mu=1}^2 D_\mu \phi D_\mu\bar{\phi} +\frac14[\phi,\,\bar\phi]^2 \biggr\},
\\
&&S_{\rm F}= \frac{1}{2g^2}\int d^2x\,\tr\biggl\{  i \sum_{\mu=1}^2 \psi_\mu D_\mu \eta + 2i \chi (D_1 \psi_2-D_2 \psi_1)  \biggr\},
\\
 && S_{\rm Y}= \frac{1}{2g^2}\int d^2x\,\tr\biggl\{ -\frac14 \eta [\phi,\eta] 
  -\chi[\phi,\chi] +\sum_{\mu=1}^2 \psi_\mu [\bar\phi,\psi_\mu]
    \biggr\}.
\eea
In contrast to the continuum theory, 
 this classification would not be strict on the lattice since the lattice action has higher order terms whose types are unclear.
At least, we will present one possibility for the lattice actions $S_{\rm lat\, B}$, $S_{\rm lat\, F}$ and $S_{\rm lat\, Y}$ such that they reproduce the continuum counterparts 
$S_{\rm B}$, $S_{\rm F}$ and $S_{\rm Y}$, respectively.

In the unimproved lattice model, we have 
\bea
&& 
S_{\rm lat\, B}=\frac{a^2}{2g^2}\sum_{x \in \Lambda_L} \tr \left\{
H^2 -\frac{i}{a^2} H \Phi_{\rm TL}+\sum_{\mu=1}^2 \nabla_\mu \phi \nabla_\mu \bar\phi +
\frac{1}{4} [\phi,\bar\phi]^2
\right\}, 
\\
&&
S_{\rm lat\, F}=\frac{a^2}{2g^2}\sum_{x \in \Lambda_L} \tr \left\{
 i \sum_{\mu=1}^2 \psi_\mu \nabla_\mu \eta +\frac{i}{a^2} \chi Q\Phi_{\rm TL} 
\right\}, 
\\
&&
S_{\rm lat\, Y}=\frac{a^2}{2g^2}\sum_{x \in \Lambda_L} \tr \left\{
-\frac14 \eta [\phi,\eta] 
  -\chi[\phi,\chi] +\frac{1}{2} \sum_{\mu=1}^2 \psi_\mu [U_\mu\bar\phi U_\mu^{-1}+\bar\phi, \psi_\mu]
\right\}.
\eea
The term $Q\Phi_{\rm TL}$ is quite complicated. Instead of giving it explicitly, let us present $Q\Phi$
since $Q\Phi_{\rm TL}(x)=Q\Phi(x)-\left\{ \frac{1}{N}\tr(Q\Phi(x)) \right\} {\bf 1}_N$. For $\Phi_1$ and $\Phi_2$, 
which are given in (\ref{Phi_admissibility}) and 
(\ref{Phi_tan2m}), respectively, $Q\Phi$ is given by
\bea
&&Q \Phi_{1}(x)= \frac{a^2}{1-\frac{1}{\epsilon^2} ||1-U_{12}(x)||^2} \nn \\
&& \hspace{1cm}
\times \bigg\{
  \Psi_{12}(x)-\Psi_{21}(x) 
	-\frac{i}{\epsilon^2} \, \tr\left\{\Psi_{12}(x) + \Psi_{21}(x) \right\} \Phi_1(x)
\bigg\}, 
\label{QPhi1}
\eea
for $||1-U_{12}(x)||<\epsilon$,
and 
\bea
&& \hspace{-1cm} Q \Phi_2(x) =\frac{2a^2}{m} (U_{12}(x)^m+U_{21}(x)^m)^{-1} \nn \\
 && \hspace{0cm}
\times 
	\sum_{k=0}^{m-1}
	\bigg\{
	U_{12}(x)^{k}\Psi_{12}(x)U_{12}(x)^{m-k-1}
	-U_{21}(x)^{k}\Psi_{21}(x)U_{21}(x)^{m-k-1} \nn \\
 && \hspace{0cm}
	-\frac{im}{2} \left(
	U_{12}(x)^{k}\Psi_{12}(x)U_{12}(x)^{m-k-1} 
	+ U_{21}(x)^{k} \Psi_{21}(x) U_{21}(x)^{m-k-1}
		\right) \Phi_2(x)
\bigg\}, 
\label{QPhi2}
\eea
where 
\be
\Psi_{\mu\nu}(x) \equiv \nabla_\mu \psi_\nu(x) U_{\mu\nu}(x) - U_{\mu\nu}(x) \nabla_\nu \psi_\mu(x) + \frac{1}{a} \left[ \psi_1(x)+\psi_2(x), U_{\mu\nu}(x) \right].
\ee
Note that $QU_{\mu\nu}(x)=i a^2 \Psi_{\mu\nu}(x)$.

For the improved lattice action, we use the variable $\psi_\mu'$ in (\ref{psi-prime}) that makes the expression as simple 
as possible. For some special cases, the action is written in an ultra-local form in terms of $\psi_\mu'$.   
The $Q^{imp}$-transformations (\ref{QU_final})-(\ref{Qchi_final}) applied to 
\eqref{Xi_lat_imp0} leads to
\bea
&& \hspace{-1cm}
S^{imp}_{\rm lat\, B}=\frac{a^2}{2g^2}\sum_{x \in \Lambda_L} \tr \left\{
H^2 -\frac{i}{a^2} H \cR_{12}\Phi_{\rm TL}+ \sum_{\mu=1}^2  \nabla_\mu \phi  \,\cS(\cR_\mu) \nabla_\mu \bar\phi +
\frac{1}{4} [\phi,\bar\phi]^2
\right\}, 
\\
&& \hspace{-1cm}
S^{imp}_{\rm lat\, F}=\frac{a^2}{2g^2}\sum_{x \in \Lambda_L} \tr \left\{
 i  \sum_{\mu=1}^2 \psi'_\mu \,\cS(\cR_\mu) \nabla_\mu \eta 
 +\frac{i}{a^2} \,\chi \,\cR_{12} \,Q^{imp}\Phi_{\rm TL} \nn \right.\\
 &&\hspace{3.8cm}
 \left.
 +\frac{i}{a^2} \chi (Q^{imp}\cR_{12}) \Phi_{\rm TL} 
\right\}, 
\\
&& \hspace{-1cm}
S^{imp}_{\rm lat\, Y}=\frac{a^2}{2g^2}\sum_{x \in \Lambda_L} \tr \left\{
-\frac14 \eta [\phi,\eta] 
  -\chi[\phi,\chi] 
  +i  \sum_{\mu=1}^2 \psi'_\mu (Q^{imp}\cS(\cR_\mu)) \nabla_\mu \bar\phi
  \right.\nn \\
&& \hspace{1.8cm} 
\left. 
+ \sum_{\mu=1}^2 \psi'_\mu[U_\mu \bar\phi U_\mu^{-1}, \cS(\cR_\mu) \psi'_\mu]
-\frac{a}{2}  \sum_{\mu=1}^2 \psi'_\mu [\cS(\cR_\mu)\nabla_\mu\bar\phi, \psi'_\mu]
  \right\},  
\eea
where 
\be
\cS(\cR) =(\cR \cR^T)^{-1},
\ee
and $Q^{imp} \Phi_{\rm TL}$ for $\Phi_i(i=1,2)$ are given by $Q \Phi$ in (\ref{QPhi1}) and (\ref{QPhi2})
with $\psi_\mu$ replaced by $\psi_\mu'$. 
The $Q^{imp}$-transformations for $\cR_\mu$ and $\cR_{12}$ remain unperformed since they are not determined in general.

The improved action becomes ultra-local for $\cR_\mu^{(-)}$ and $\cR_{12}^{(-)}$ given in  (\ref{Rmu-_r}) and (\ref{R12-}).
In the case of 
\bea
&&\left. \cR^{(-)}_\mu \right|_{r=\frac{1}{2}} = \left(1-\frac{a}{2}\nabla_\mu\right)^{-1},\\
&&\left. \cR^{(-)}_{12} \right|_{r=\frac{1}{2}} = 1- \frac{a}{2} \sum_{\mu=1}^2 \nabla_\mu,
\eea
we find that
\bea
&& 
S^{imp}_{\rm lat\, B}=\frac{a^2}{2g^2}\sum_{x \in \Lambda_L} \tr \left\{
H^2 -\frac{i}{a^2} H \left(2\Phi_{\rm TL}-\frac{1}{2}\sum_{\mu=1}^2U_\mu \Phi_{\rm TL}U_\mu^{-1} \right)
\right.\nn\\
&&
\hspace{3.8cm}
 \left.
+ \frac{1}{4} \sum_{\mu=1}^2 \left(3\nabla_\mu \phi -\nabla_\mu^*\phi \right) \left(3\nabla_\mu \bar\phi -\nabla_\mu^*\bar\phi \right) +
\frac{1}{4} [\phi,\bar\phi]^2
\right\}, 
\\
&&
S^{imp}_{\rm lat\, F}=\frac{a^2}{2g^2}\sum_{x \in \Lambda_L} \tr \left\{
 i \sum_{\mu=1}^2 \psi'_\mu\left(1-\frac{3a^2}{4}\nabla_\mu\nabla_\mu^* \right) \nabla_\mu \eta 
 +\frac{i}{a^2} \chi \left(1-\frac{a}{2}\sum_{\mu=1}^2\nabla_\mu \right) Q^{imp}\Phi_{\rm TL} \nn \right.\\
 &&\hspace{3.8cm}
 \left.
 -\frac{1}{2a} \sum_{\mu=1}^2 \chi \left[U_\mu \Phi_{\rm TL}U_\mu^{-1}, \psi_\mu' \right] 
\right\}, 
\\
&&
S^{imp}_{\rm lat\, Y}=\frac{a^2}{2g^2}\sum_{x \in \Lambda_L} \tr \left\{
-\frac14 \eta [\phi,\eta] 
  -\chi[\phi,\chi] 
  \right.\nn \\
&& \hspace{3.8cm} 
+\frac{1}{8}
\sum_{\mu=1}^2
  \psi_\mu' \left[
       13 \bar\phi +13 U_\mu \bar\phi U_\mu^{-1} -3 U_\mu^{-1} \bar\phi U_\mu -3 U_\mu U_\mu \bar\phi U_\mu^{-1} U_\mu^{-1}, \psi_\mu' 
\right] \nn \\
&& \hspace{3.8cm} 
-\frac{3}{4}
\sum_{\mu=1}^2
  \psi_\mu' 
 [ U_\mu \bar\phi U_\mu^{-1}+ U_\mu^{-1}\bar\phi U_\mu, U_\mu^{-1} \psi_\mu' U_\mu]
\bigg\},  
\eea
where $Q^{imp} \Phi_{\rm TL}$ for $\Phi_i(i=1,2)$ are again given by $Q \Phi$ in (\ref{QPhi1}) and (\ref{QPhi2})
with the replacement of $\psi_\mu$ by $\psi_\mu'$. 
These actions are clearly ultra-local and suitable for numerical simulations.

We can show that the Yukawa interactions of the improved lattice action coincide with those of the unimproved one
for  (\ref{Rmue})
since $\cS(\cR^{(\e)}_\mu)=1$.  Then we can show that
\bea
&& 
S^{imp}_{\rm lat\, B}=S_{\rm lat\, B}|_{\Phi \rightarrow \Phi^{imp}},
\\
&&
S^{imp}_{\rm lat\, F}=S_{\rm lat\, F}|_{\psi_\mu\rightarrow \psi_\mu', \,Q\Phi \rightarrow Q^{imp}\Phi^{imp}},
\\
&&
S^{imp}_{\rm lat\, Y}=S_{\rm lat\, Y}|_{\psi_\mu \rightarrow \psi_\mu'},
\eea
where 
\be
\Phi^{imp} = \cR_{12} \Phi.
\ee
The difference is only from the definition of the improved lattice field tensor $\Phi^{imp}$ given via $\cR_{12}$.   
Furthermore, if we use  (\ref{R12e}) for $\cR_{12}$ and integrate the auxiliary field $H$, 
the factor $\cR_{12}$ disappears in the boson action since $\cS(\cR^{(e)}_{12})=1$. 
In other words, the improved actions are the same with those of the unimproved theory for the boson and the Yukawa interactions.
The difference remains only in the fermion kinetic term relevant to $\chi$.
This would have some theoretical importance and should be studied further.

\section{Evaluation of the determinant in (\ref{det_D_solution})}
\label{app:determinant}

In this appendix, we calculate the determinant of the $L\times L$ circulant matrix, 
\begin{equation}
R_L=
\begin{pmatrix} A & B&   &  C \\
C & \ddots &  \ddots&     \\
   & \ddots & \ddots  &  B  \\
B  &            &   C  & A    
\end{pmatrix}, 
\end{equation}
for constant $A$, $B$ and $C$. 
To this end, it is convenient to introduce a purely tridiagonal matrix: 
\be
Q_L\equiv \begin{pmatrix} A & B&   &   \\
C & \ddots &  \ddots&     \\
  & \ddots & \ddots  &  B  \\
  &            &   C  & A    \end{pmatrix}. 
\ee

In computing $\det R_L$, the cofactor expansion with respect to the first and second rows or columns gives 
\be
\det R_L = A\det Q_{L-1} -2BC\det Q_{L-2} -(-B)^L -(-C)^L.
\label{detRLapp}
\ee
Similarly, 
\be
\det Q_L = A\det Q_{L-1} -BC\det Q_{L-2}. 
\label{detQL}
\ee
Defining the solutions of the quadratic equation $x^2-Ax+BC=0$ as 
\be
\xi_\pm \equiv \frac12\left(A\pm \sqrt{A^2-4BC}\right), 
\ee
the recursion equation (\ref{detQL}) is solved as 
\begin{equation}
	\det Q_L = 
	\begin{cases}
		(L+1) \xi_+^L & \  \   {\rm if} \ A^2=4BC  \\
		\frac{\xi_+^{L+1}-\xi_-^{L+1}}{\xi_+-\xi_-} &   \ \ {\rm otherwise},
	\end{cases}
\end{equation}
Plugging this to (\ref{detRLapp}) leads to the simple expression
\be
\det R_L = \xi_+^L +\xi_-^L -(-B)^L-(-C)^L. 
\label{detRL_solapp}
\ee

\providecommand{\href}[2]{#2}\begingroup\raggedright\endgroup

\end{document}